\documentclass[11pt,a4paper]{article}
\usepackage[english]{babel}
\usepackage[utf8x]{inputenc}
\usepackage[T1]{fontenc}
\usepackage{mwe}
\usepackage[a4paper,top=3cm,bottom=2cm,left=3cm,right=3cm,marginparwidth=2cm]{geometry}
\expandafter\let\csname equation*\endcsname\relax
\expandafter\let\csname endequation*\endcsname\relax
\usepackage{amsmath}
\usepackage{graphicx}
\usepackage{lipsum}
\usepackage[colorinlistoftodos]{todonotes}
\usepackage[colorlinks=true, allcolors=blue]{hyperref}
\usepackage{caption}
\usepackage{subcaption}
\usepackage{authblk}
\usepackage{float}
\usepackage[section]{placeins}
\usepackage{cite}
\usepackage{booktabs}

\begin{document}
\title{ NV$^-$ with nitrogen}
\author[1]{Neil B. Manson\footnote{Corresponding author: Neil.Manson@anu.edu.au}${}^{,}$}
\author[2]{Michael S. J. Barson}
\author[1]{Morgan Hedges}
\author[1]{Priya Singh}
\author[1]{Sophie Stearn}
\author[1]{YunHeng Chen}
\author[1]{Lachlan Oberg}
\author[3]{Carlos A. Meriles}
\author[1]{Marcus W. Doherty}
\affil[1]{Department of Quantum Science and Technology, Research School of Physics, Australian National University, Canberra, A.C.T. 2601, Australia.}
\affil[2]{School of Physics and Astronomy, Monash University, Melbourne, Vic, Australia}
\affil[3]{Department of Physics, CUNY - City College of New York, New York, New York 10031, USA}
\maketitle

\section{Abstract}

Single NV$^-$ centres in diamond have a spin ground state that can be polarised and read optically and these properties have led to many applications. Ensembles of NV$^-$ centres in bulk diamond have similar properties and are likewise being adopted for applications. However, in bulk diamond there is a degrading of the properties of NV$^-$ compared to single centres. This is known and occasionally mentioned in the literature but the processes that give rise to the less desirable properties have not been presented in any detail. Here NV-diamonds containing a concentration of nitrogen from 1 ppm to 600 ppm are investigated and indeed found the higher the nitrogen concentration the more is the degrading. The emission contrast (between polarised and not polarised) in both the visible and infrared decreases with nitrogen concentration. The lifetimes shorten with nitrogen concentration and in addition there is a change of the amplitude of the the NV charge state via two mechanisms, change of the optical cycle, and there are some optical transitions that do not conserve spin. Each of factor will be illustrated and discussed.

\section{Introduction}
There is considerable interest in the NV$^-$ defect centre in diamond as it has a spin ground state that can be initialised and readout optically \cite{Doherty2013}. The defect centre can be detected at the single site level and with the spin capabilities the properties have led to numerous applications including detection of magnetic fields \cite{Balasubramanian2008,Maze2008,Tetienne2013}, electric fields\cite{Dolde2011,Barson2021,McCloskey2022}, pressure \cite{Doherty2014} and temperature \cite{Acosta2010} and also for development of room-temperature quantum computers and quantum sensing \cite{Childress2013,Degen2017}. As well as single NV$^-$ centers there is interest in ensembles of NV$^-$ centres for applications because with multiple centres larger signals and better signal-to-noise can be expected. This maybe the case but what is not always recognised is that the properties of the NV$^-$ centres within diamond containing nitrogen can be significantly degraded in comparison to the isolated  NV$^-$ case. To obtain the nitrogen to form the centres plus electron donors it is common to fabricate samples starting with 1b diamonds with significant nitrogen concentrations. It is the presence of the nitrogen that is not specifically involved with the centres (the nitrogen bath) that results in the less desirable properties. 

There are publications reporting on the significance of nitrogen in diamond such as the review by Ashfold \textit{et al.} \cite{Ashfold2020} and there are topics very relevant to the NV$^-$ centre such as the ground state spin coherence dependence on nitrogen \cite{Bauch2020}. There are also other studies of the degrading of certain properties of the NV$^-$ centre due to nitrogen \cite{Capelli2022}. The present work restricts the attention to the optical properties and the optical cycle where there is a difference from single site NV properties that result from the presence of nitrogen. The information is relevant to obtaining NV samples with good spin polarisation and high quantum efficiency. There are four factors that will be considered. (i) One is the change of charge state that occurs with optical excitation of NV$^-$. This gives rise to reduced emission with a shorting of the lifetime and an effective reduction of the NV$^-$ concentration. Also $\text{NV}^- \leftrightarrow \text{NV}^0$ charge cycling reduces the spin polarisation. (ii) There is a second mechanism for charge transfer due to the ionisating the nitrogen bath. This ionisation can be significant for energies greater than 2.2 eV and so excitation lasers such as that at 532 nm or 520 nm with energies > 2.2 eV can introduce a density of N$^+$ ions. The subsequent relaxation of these ions can cause a change of the NV charge state. This provides a second $\text{NV}^- \leftrightarrow \text{NV}^0$ cycling that again reduces spin polarisation.	(iii) Another factor is that the intersystem crossing giving rise to spin polarisation is modified compared to single sites. There is not a great loss of spin polarisation but significant shortening of lifetime leading again to the centres being `less bright'.  (iv) A further deteriorating factor is the occurrence of non-spin-conserving transitions. As these transitions were observed in samples that contained nitrogen \cite{Reddy1987} comments of the situation is merited.

The variation of the properties with the nitrogen concentration are reported for concentration of 1 ppm to 600 ppm nitrogen. It would be desirable to model the behavior in each case although it is recognized that the properties of each NV within an ensemble depends on its local environment. For example there is a dependence on the distance to nearest nitrogen donor and this can vary with optical excitation \cite{Manson2006}. The significance is that for a given sample with known nitrogen concentration there is not a fixed set of parameters but rather a distribution of parameters and the distribution can be modified by experimental conditions. Simple rate equation modelling is adopted here and does provide some insight into observation and trends. However, with ensembles the calculation with such a set of fixed parameters do not and cannot be expected to give exact agreement with observations. 
	
\section{NV charge state}
	
\begin{figure}[h!]
\centering
\includegraphics[width=1.0\textwidth]{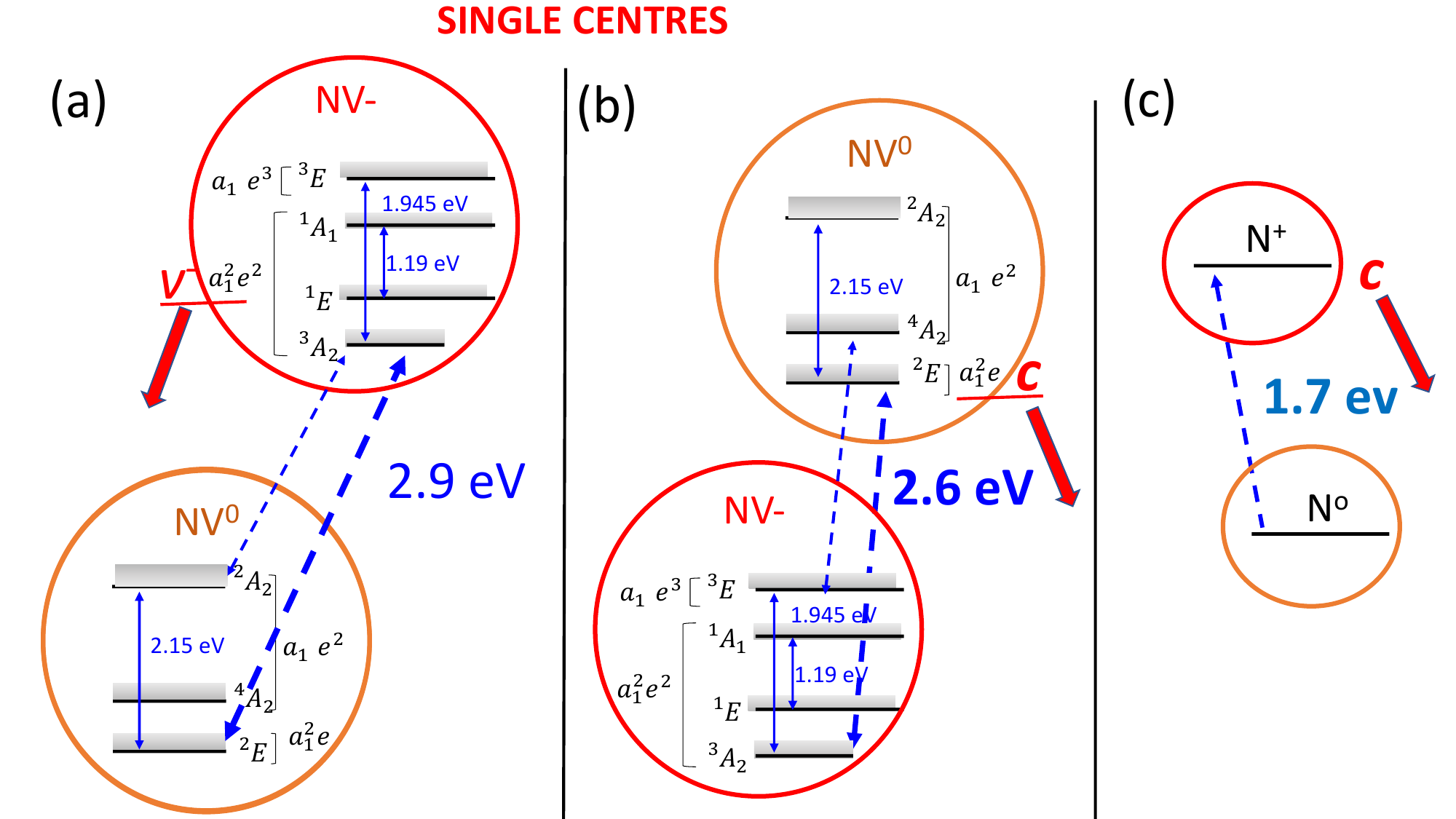}
\caption{\label{single} (a and b) The figures gives the relative energies of NV$^-$ and NV$^0$ single centres that can occur when there is little nitrogen in the diamond lattice. Transition energies that can result in $\text{NV}^0 \leftrightarrow \text{NV}^-$ charge transfer are indicated \cite{Razinkovas2021}. For these direct transitions given by bold dashed arrows (one photon) a hole V$^-$ is created in the valence band or an electron C in the conduction band. There are also transitions from other than the ground state and these involve two photons, and two examples are indicated by light arrows.  (c) Should there be single substitutional nitrogen in the lattice it can be ionized N to N$^+$ with an energy of 1.7 eV and an electron excited to the conduction band.}
\end{figure} 
	
\begin{figure}[ht!]
\centering
\includegraphics[width=1.0\textwidth]{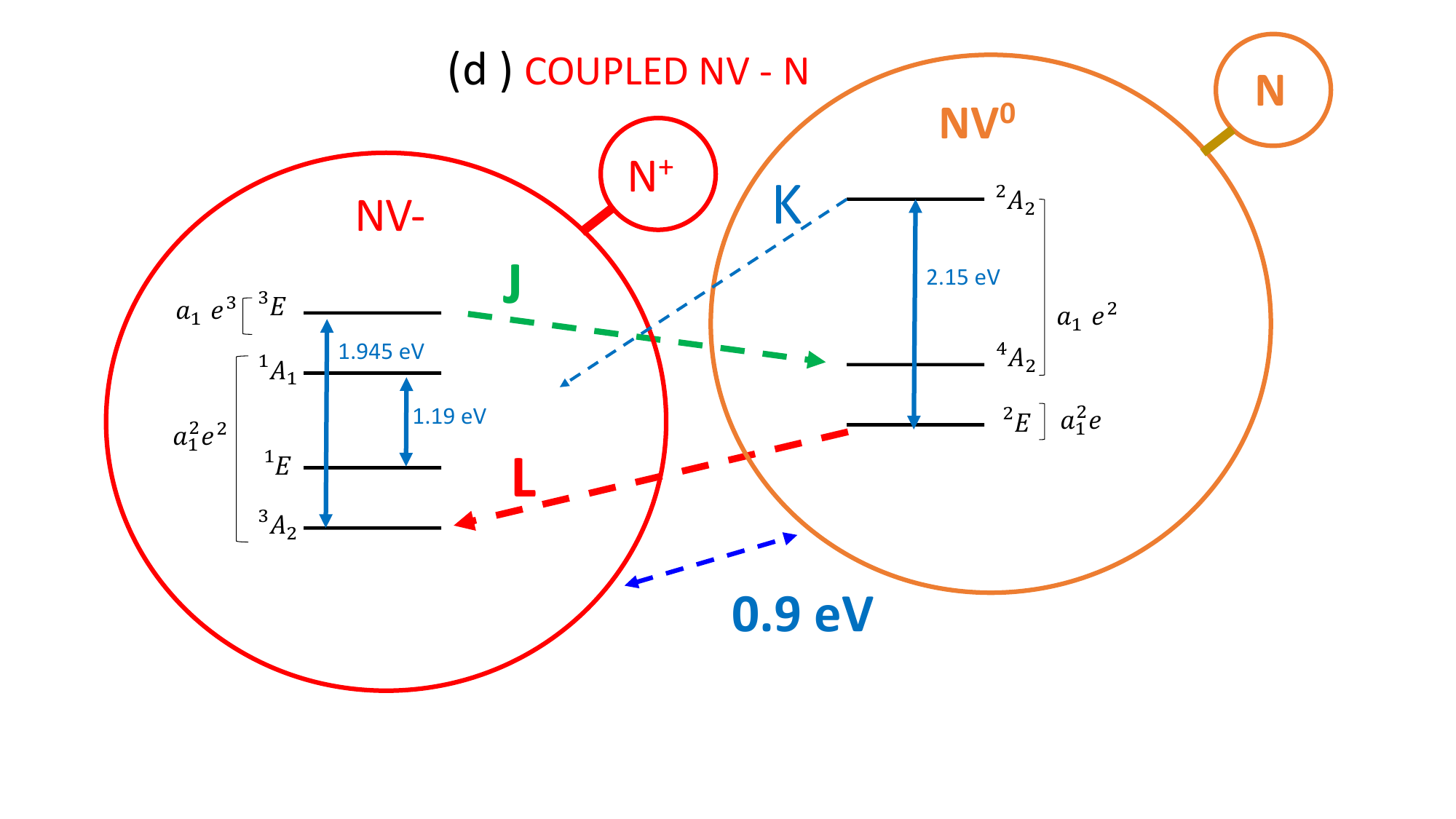}		
\caption{\label{pair}The figure indicates the situation when there is significant concentration of nitrogen in the diamond with some of the nitrogen atoms adjacent to the NV centre. From the energies in Fig \ref{single} it can be seen that the energies between the pairs NV$^-$/N$^+$ and NV$^0$/N$^0$ are greatly reduced from the isolated NV case and the separation is of the order of 0.9 eV. There can be an effective coupling between NV and adjacent N and tunnelling of an electron between the NV and N with no involvement of conduction-band electrons or valence-band holes. The line joining NV and N are indicates the coupling and the possible tunnelling and there can be some overlap of wavefunctions. There can be various distances between the NV and N and the tunnelling rates or transition probabilities will vary exponentially with the separation. With the energies indicated there can be various transitions that are energy favourable and can occur via spontaneous non-radiative relaxation. The tunnelling will involve a change of only one molecular orbital this factor restricts the final state. Some of the spontaneous transitions are given by dashed arrows. For example from the ground state of NV$^0$ to NV$^-$ (red dashed arrow, L), from the excited state of NV$^0$ to NV$^-$ (blue dashed arrow, K) and also from the excited state of NV$-$ to NV$^0$ (green dashed arrow J). These transitions are discussed in the text.}
\end{figure}
	
Before dealing with the specifics of change of charge state it is appropriate to introduce the energy levels of NV systems. The properties of vacancy-related defects in diamond are frequently treated in relation to one electron states that occur in the gap between the valence and conduction band. In T$_d$ symmetry (the carbon vacancy) the molecular orbitals that can be formed from the four carbon dangling bonds give two states: one of a$_1$ symmetry and one three-fold degenerate of t$_2$ symmetry. When one of the carbon atoms is replaced with a nitrogen atom to give the NV centre the symmetry is lowered to C$_{3v}$ and the t$_2$ states splits into a second non-degenerate a$_1$ state and a two-fold degenerate e state. There are two a$_1$ states but one is always occupied lying in the valence band and can be neglected. Hence for NV there are one a$_1$ state and a degenerate e state within the band gap between valence and conduction band. A more complete treatment of the molecular orbital treatment is given in reference \cite{Doherty2011}.
	
The energy of these latter one-electron states can be drawn in relation to the valence and conduction bands. However, when these states are occupied by the three electrons for NV$^0$ or four for NV$^-$ it can be misleading to include the energy levels of the \textit{multi-electron} states in relation to the $one-electron$ valence and conduction band states.  However, when the one electron states are filled they give the NV$^0$ and NV$^-$ energies as given in Fig \ref{single}. Both charge states can exist or co-exist in a diamond crystal. There can also be charge transfer between the two states and the one-photon transitions that change the charge are given in Fig \ref{single}. To gain an electron as required for NV$^0$ $\rightarrow$ NV$^-$ an electron is excited out of the valence band to a molecular orbital state and subsequently there is a relaxation to fill the hole in the valence band (Fig \ref{single}(a)). To decrease the number from 4 to 3 for NV$^-$ $\rightarrow$ NV$^0$ an electron is excited out of one of the one-electron states to the conduction band followed by the relaxation of the conduction band electron (Fig \ref{single}(b)). There can be other transitions that change charge but these occur from other that the ground state and involve two-photon excitation. One photon to populate the excited state followed by the charge transfer transition which again must involve the valence or conduction bands. Therefore, for single centres the specific charge transfer transition always creates an electron in the conduction band (given by C) or a hole in the valence band (given by V). The conduction electron or the valence hole will subsequently relax leaving the centre with the altered charge state.
	
When there is a concentration of nitrogen in the diamond the charge transfer is different from that described above for single centres. In Fig \ref{single}(c) shows that substitutional nitrogen in diamond can also be ionised to the conduction band to give a positely charged N$^+$ requiring an energy of >1.7 eV. The energies as given in Fig \ref{single} indicates that an NV$^-$ together with N$^+$ have an energy compared to NV$^0$  combined with N that differs by only 0.9 eV. The energy level scheme is given in Fig \ref{pair} and there is no involvement of valence or conduction band.  The NV and N pair could be separated by many nm's but there can be electron tunnelling between the NV and the N and the probability of this tunnelling will vary exponentially with NV and N separation\cite {Alkauskas2023}. Should there be a NV$^0$ and an adjacent N it is noted that the pair has higher energy than NV$^-$ and N$^+$. A transfer of an electron from the N to NV$^0$ is likely to occur although if separation of NV from N is large the rate may be slow (or inhibited). Such a transition is given by the red-dashed arrow L in Fig \ref{pair}. The energy mismatch will involve creation of phonons. Other energetically favourable transitions J and K  are also indicated. These are from excited states and so will involve one photon to access the excited state, but the transfer will be spontaneous. This diagram Fig \ref{pair} will be utilised to explain many of the processes in the following sections.
	
\subsection{NV charge state -- Optical spectra} 
Measurements have been made of samples that contain a concentration of substitutional nitrogen impurities. The crystals, mainly HPHT, were acquired over years from various sources. The samples were electron irradiated and annealed to produce the NV defects. Details are not given as the topic has been covered by others including a  discussion in reference  \cite{Ishii2022}. The concentrations have been established from optical, ESR and FTIR measurements and the NV concentrations of NV is low of about 1 ppm such that there is negligible NV--NV interactions. 
	
The assertion is that the properties vary as a consequence of the local environment and in particular the distribution of nitrogen. It should not depend to first order as to how the crystals are grown only as preparation techniques may vary the local environment. Of course the growth and subsequent treatment of the diamonds incorporating NV centres is essential and there are many reports covering this important topic \cite{Achard2020,Edmonds2020}.  However, the present work is mostly relevant to bulk diamond and not to the situation for NV adjacent to surfaces or for nanodiamonds (although only because the local environment is very different). It is also worth noting that the measurements are largely CW (or very low frequency modulation) and although intensities can seem `high' they are not at intensities attainable with pulsed lasers. Single centre processes such as the two-photon charge transfer mentioned earlier are essentially `intrinsic' and so they are still possible. These \textit{intrinsic} processes may occur in present samples at high intensities but in the present spectroscopic measurements these are not observed as the alternative processes as a consequence of the presence of nitrogen dominates. The situation is very different for detection techniques that are sensitive to charge movement within conduction or valence bands such as photoconductivity. In this case the two-photon ionisation will give responses and there is very significant work using photoelectron detection of the spin state of the NV centre. A excellent review of the situation is given by Bourgeois \textit{et al.} \cite{Bourgeois2020} but electron detection is not involved here.
	
\begin{figure}[t!]
\centering
\includegraphics[width=1.0\textwidth]{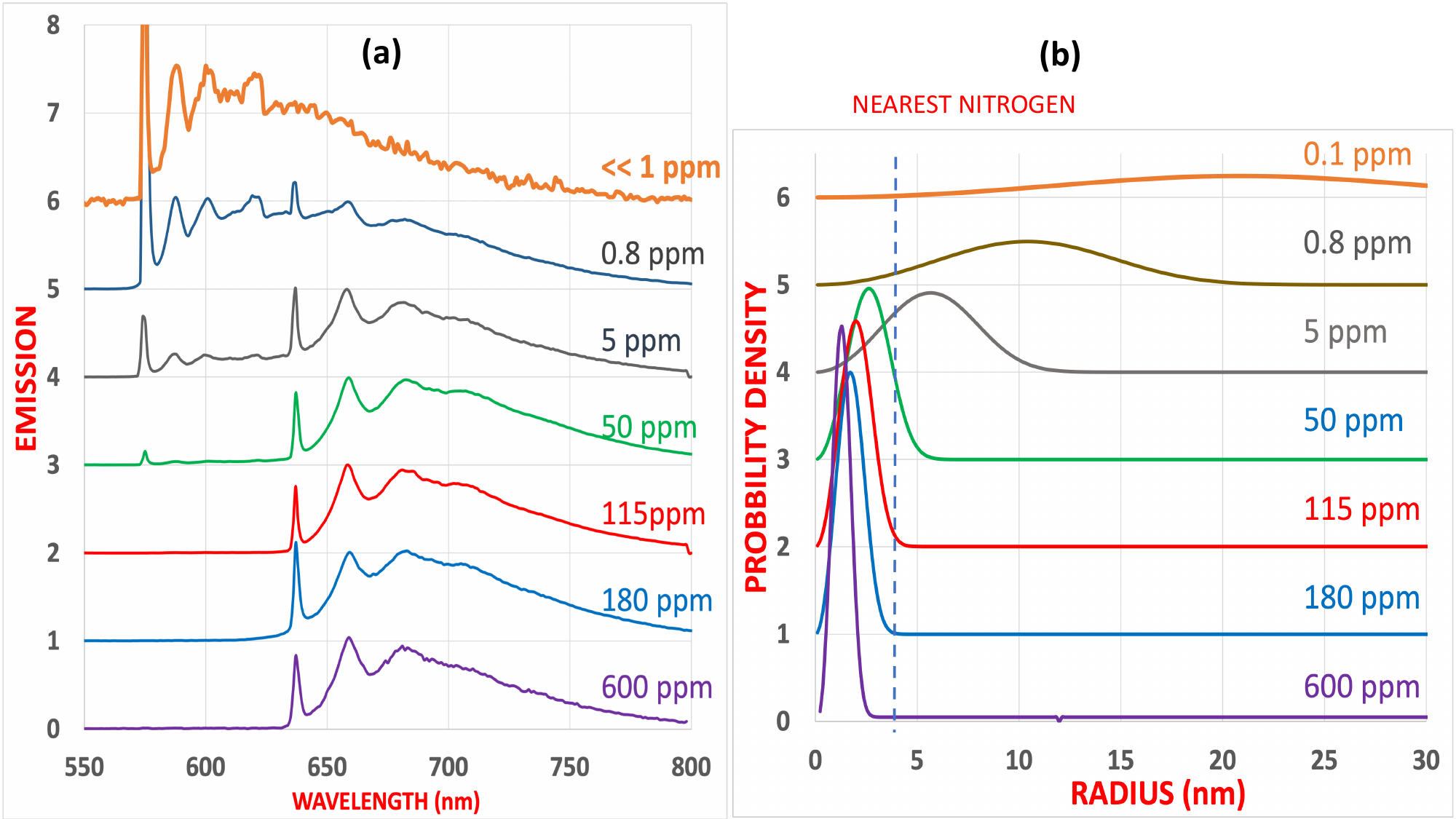}
\caption{\label{samples}  (a) Emission of samples at 77 K using low intensity 532 nm excitation where the intensities do not alter the NV charge state. The emission in the spectral range between 575 nm and 637 nm is due to NV$^0$ whereas to the long wavelength side of 637 nm the emission is mainly due to NV$^-$.  (b) Gives the distribution of nearest nitrogen atoms for the corresponding concentration of N impurities. The dashed line at 4 nm indicates a distance for which there is a mark change in the relative concentration of NV$^-$ and NV$^0$ centres. }
\end{figure}

\subsection{ NV charge state:  Tunneling transition L }

A large number ( > 30) of samples have been investigated exhibiting distinct trends with nitrogen concentrations and these trends are illustrated experimentally using a smaller number of representative samples. For example, emission for seven samples using low excitation (at least an order of magnitude lower than that giving any change of charge state) are shown in Fig \ref{samples}. For high nitrogen concentrations the emission is entirely that of NV$^-$ whereas there is an increasing NV$^0$ contribution as the nitrogen concentration is reduced. This particular trend is not new, reported by many and is described by Collins \cite{Collins2002} as a consequence of a `local Fermi level'. There is also a discussion of the equilibrium situation given in \cite{Shinei2021}. In Fig \ref{samples} (b) the density of the nearest nitrogen impurities are calculated for the various nitrogen concentrations. The figure shows the high density of close nitrogen atoms at the high nitrogen concentrations and the wide distribution for the low concentrations. For high concentrations of $>$100 ppm nearly all NV centres have nitrogen atoms within 4 nm and this correlates with the samples where electrons have tunnelled from adjacent nitrogen to NV to give centres that are entirely in the negative charge state. For lower N concentrations (higher in Fig \ref{samples}) there are a significant fraction of the centres with the nearest neighbour nitrogen atoms at distances larger than 4 nm. For these larger distances tunnelling time becomes prohibited and correspondingly the centres remain neutral and NV$^0$ emission is obtained. With the exponential variation with distance, rates might vary from pico-seconds at say 1 nm distance to many thousands of sec at 5 nm. The transition involved that give the spectra in Fig \ref{samples} is that given by the red dashed arrow L in Fig \ref{pair}. As there is a gain in energy the transitions are spontaneous and can and have occurred in this case in the dark.

\subsection{NV charge state: Tunneling transition K}
\begin{figure}[ht!]
  \centering
    \begin{subfigure}[b]{0.315\textwidth}
        \includegraphics[width=\textwidth]{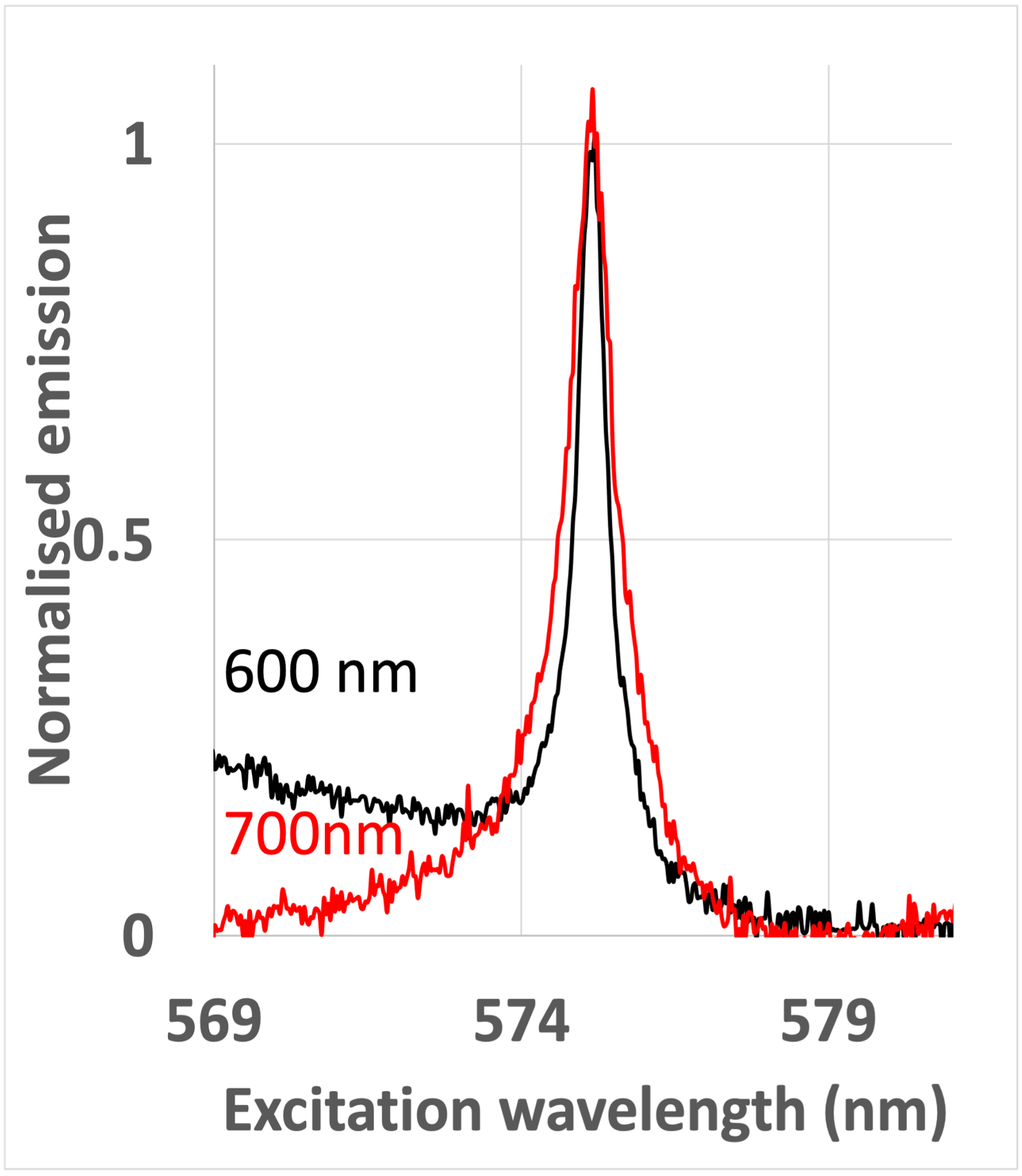}
        \caption{}
        \label{}
    \end{subfigure}
~   
    \begin{subfigure}[b]{0.327\textwidth}
        \includegraphics[width=\textwidth]{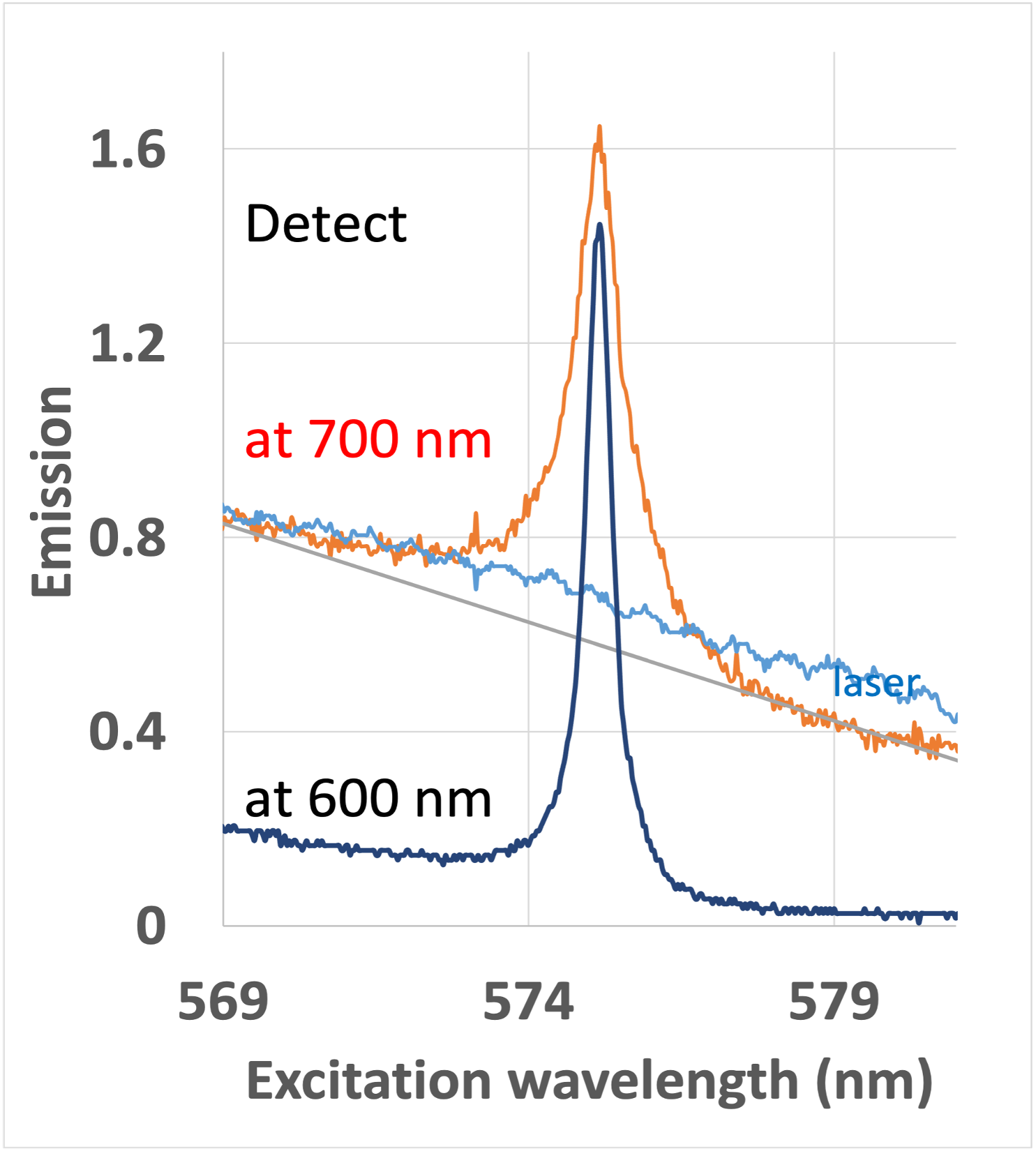}
        \caption{}
        \label{}
    \end{subfigure}
~
    \begin{subfigure}[b]{0.31\textwidth}
        \includegraphics[width=\textwidth]{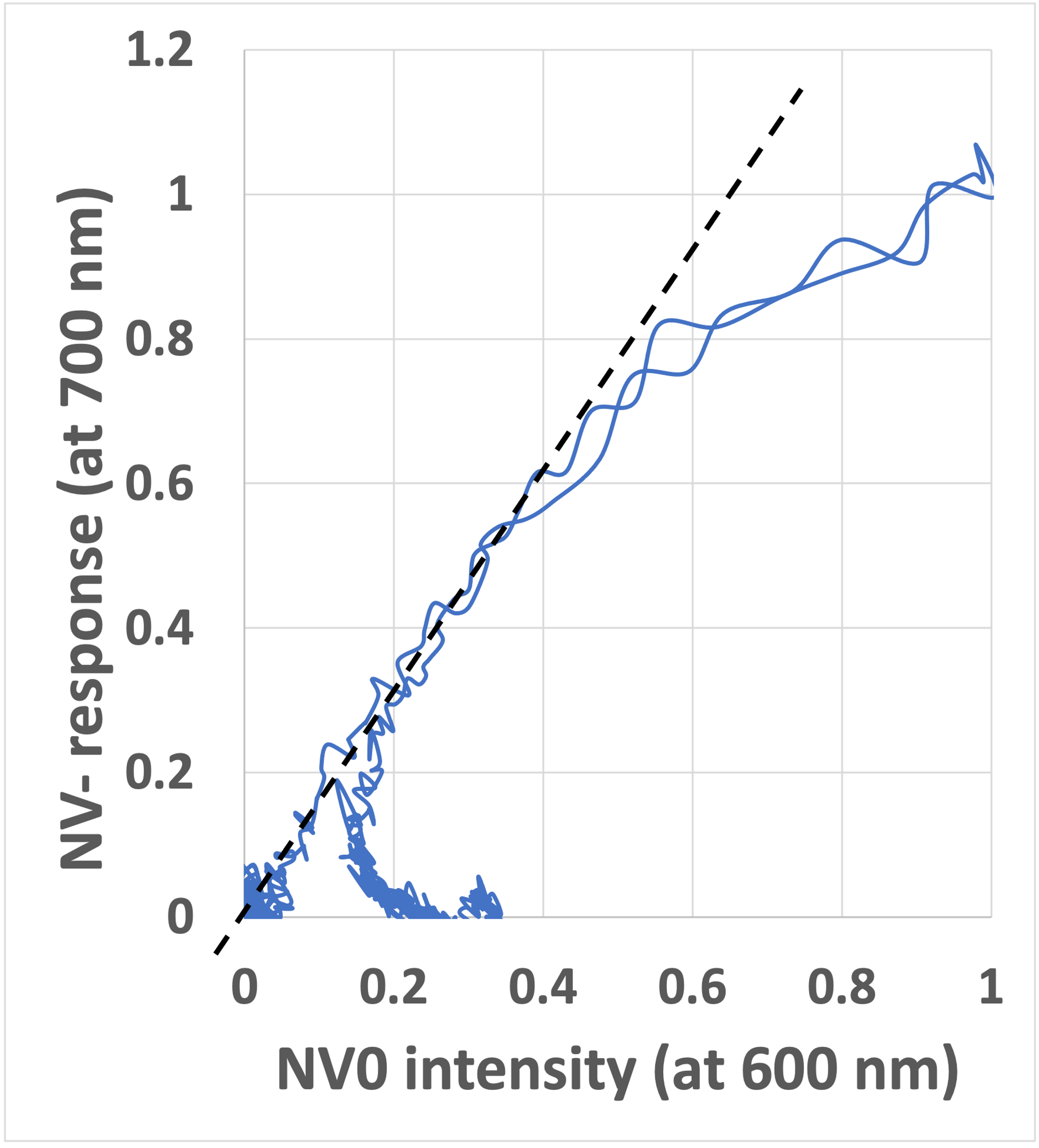}
        \caption{}
        \label{}
    \end{subfigure}
\caption{\label{nv0 excit} Experiments to illustrate that the NV$^0$ -- NV$^-$ charge transfer from excited state of NV$^0$ is linear (transition K in Fig \ref{pair}).  Traces are for 20 ppm N sample and excitation of NV$^0$ is found to increase NV$^-$ emission. When sweeping the excitation wavelength across the NV$^0$ zero-phonon energy as in (a) it is seen that there is an increase at 700 nm (ie NV$^-$ emission) whereas detection at 600 nm gives the strength of the NV$^0$ excitation. The traces in (b) are as in (a) but with background subtracted (c) Gives the increase of  NV$^-$ plotted as a function of the NV$^0$ excitation and shows the response is linear although saturates at the highest excitation values. It is accepted that transition L will also be involved but is constant over the narrow spectral range.}
\end{figure}

There is a second transition involving a charge transfer from NV$^0$ to NV$^-$ (NV$^0$ -- N to NV$^-$ -- N$^+$) in this case from the excited state $^2$A$_2$ of NV$^0$ to NV$^-$ as given by the blue dashed arrow K in Fig \ref{pair}. By exciting the zero-phonon line of NV$^0$ at 575 nm there is an increase in the NV$^-$ intensity with an electron from an adjacent N. Plotting the intensity of NV$^0$ versus this excitation it is found the increase is linear Fig \ref{nv0 excit} (c). So for this charge transfer from NV$^0$ requires one photon to access the excited state but the charge transfer is spontaneous. As the charge transfer can be observed for centres where in the dark the centres remain in the NV$^0$ ground state, this indicates that the transition probability of K is greater than that for L.

\begin{figure}[ht!]
\centering
\begin{subfigure}[b]{0.315\textwidth}\includegraphics[width=\textwidth]{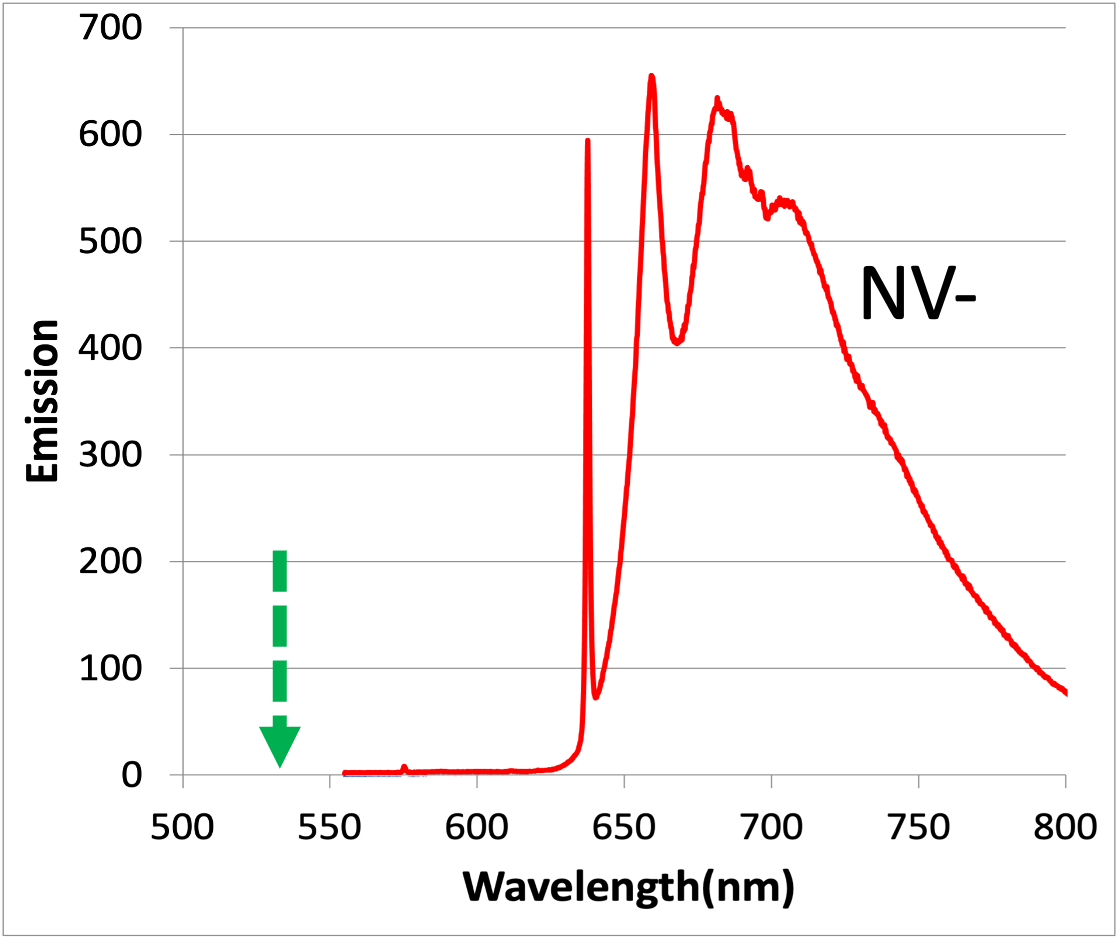}\caption{}\end{subfigure}
\begin{subfigure}[b]{0.312\textwidth}\includegraphics[width=\textwidth]{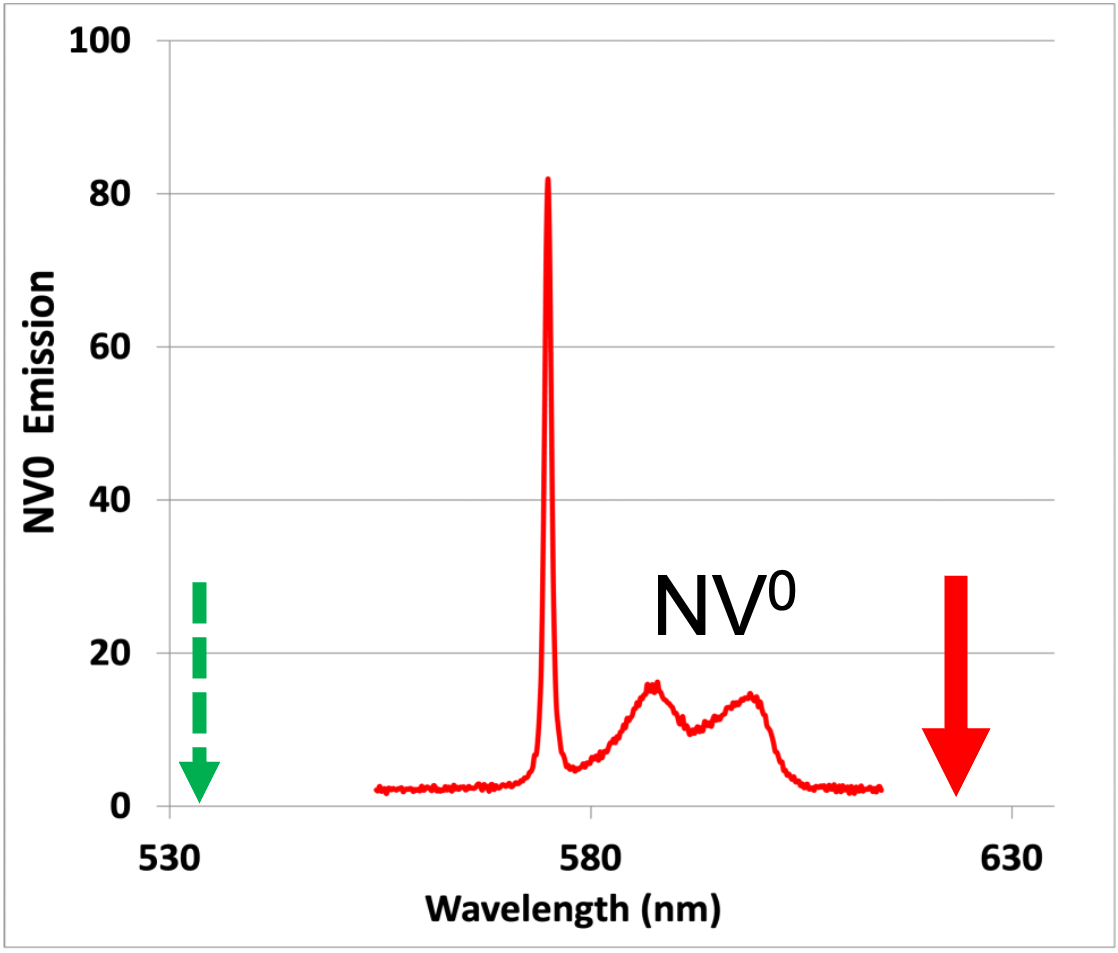}\caption{}\end{subfigure}
\begin{subfigure}[b]{0.35\textwidth}\includegraphics[width=\textwidth]{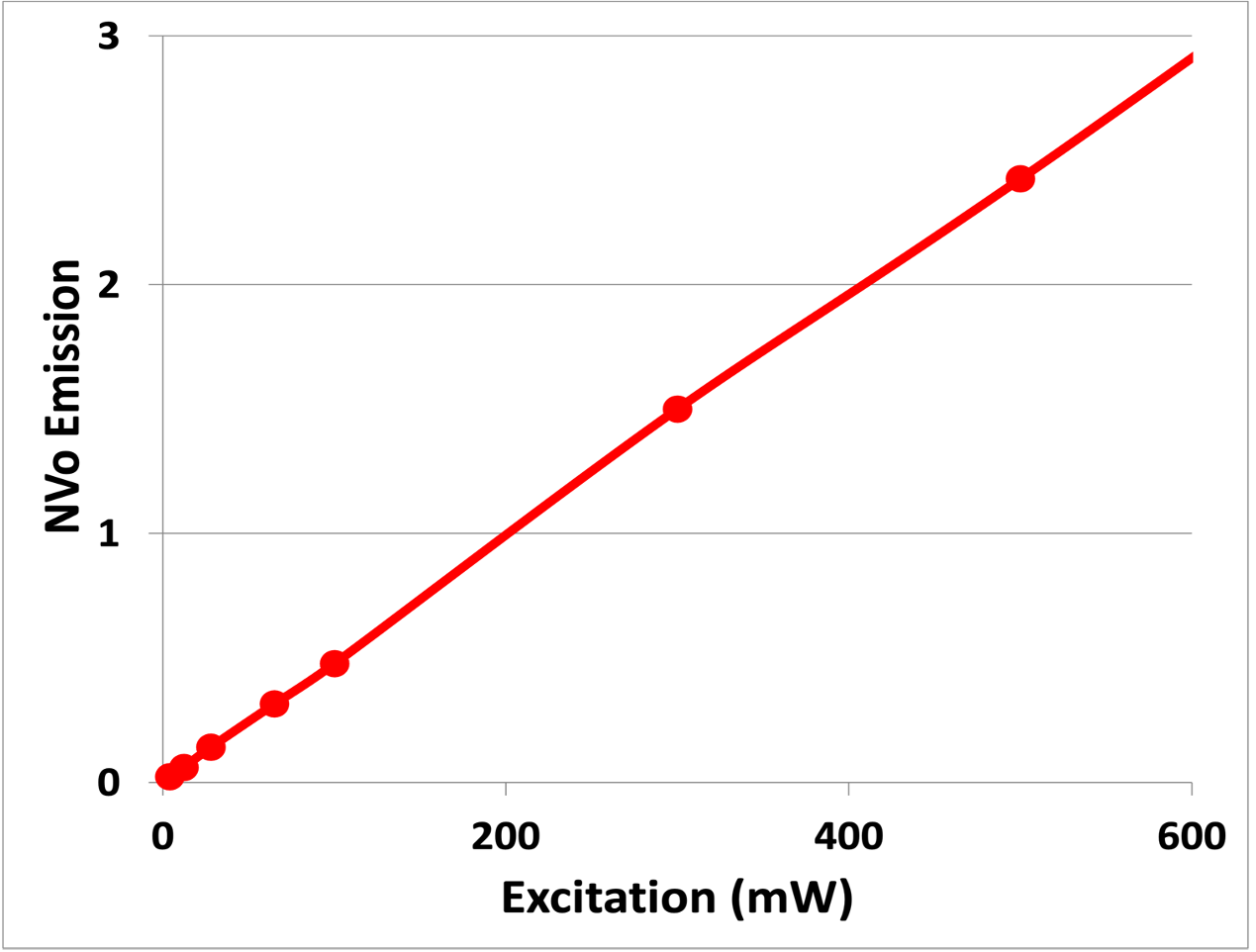}\caption{}\end{subfigure}
\caption*{1. NV${}^0$ created by 632~nm excitation and detected using weak 532~nm.}
\begin{subfigure}[b]{0.355\textwidth}\includegraphics[width=\textwidth]{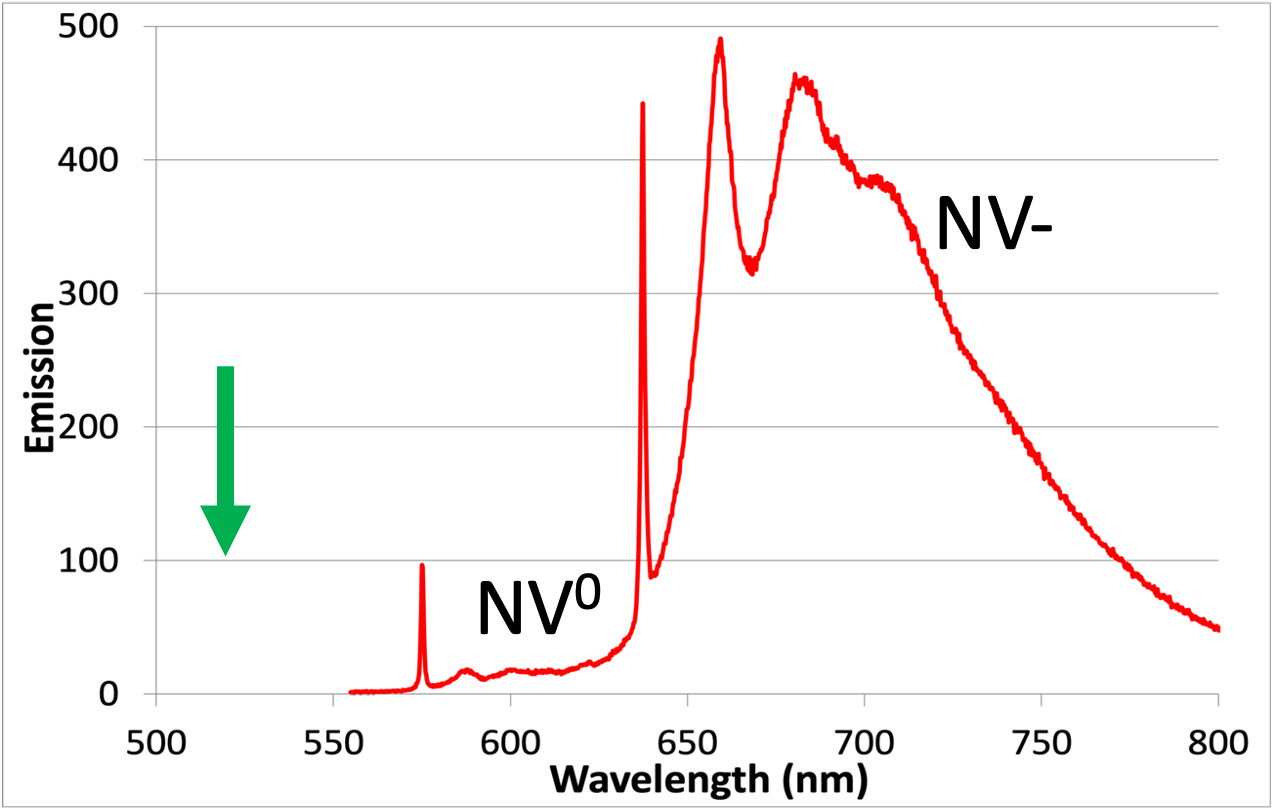} \renewcommand{\thesubfigure}{a}\caption{}\end{subfigure}
\begin{subfigure}[b]{0.321\textwidth}\includegraphics[width=\textwidth]{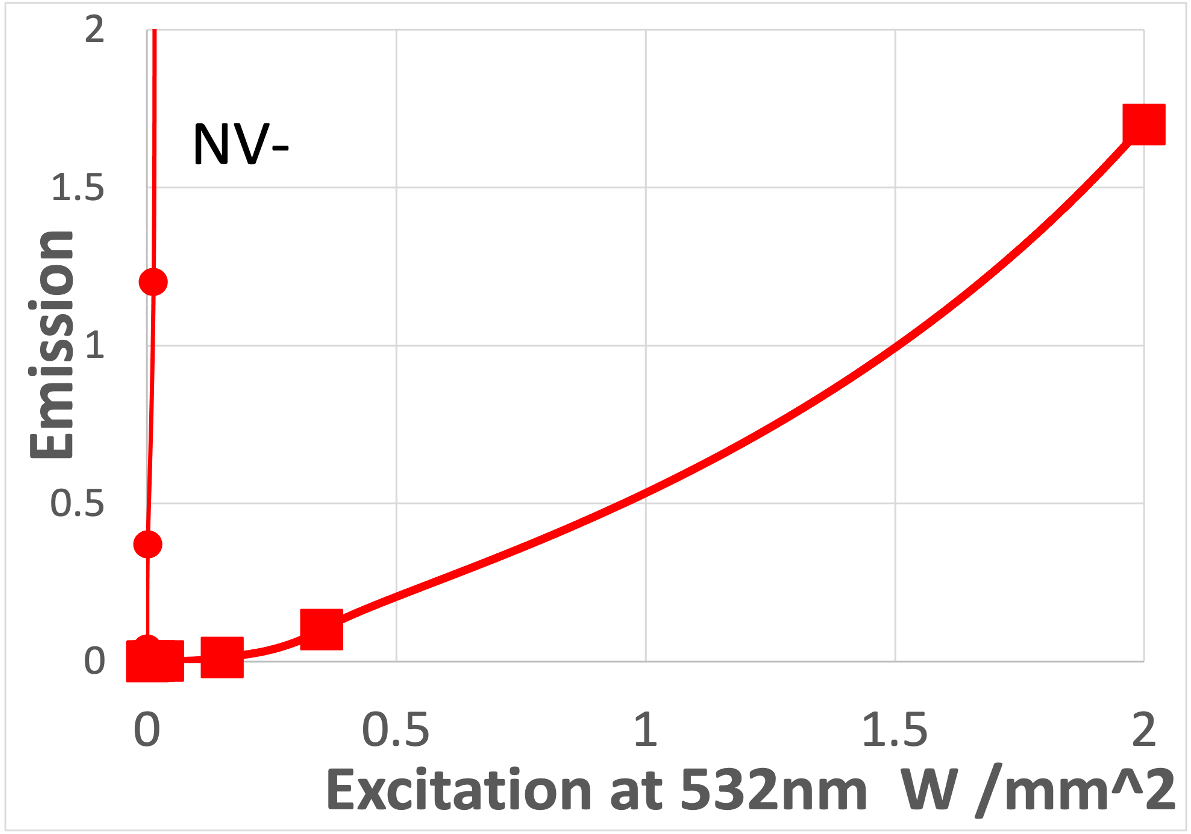} \renewcommand{\thesubfigure}{b}\caption{}\end{subfigure}
\begin{subfigure}[b]{0.275\textwidth}\includegraphics[width=\textwidth]{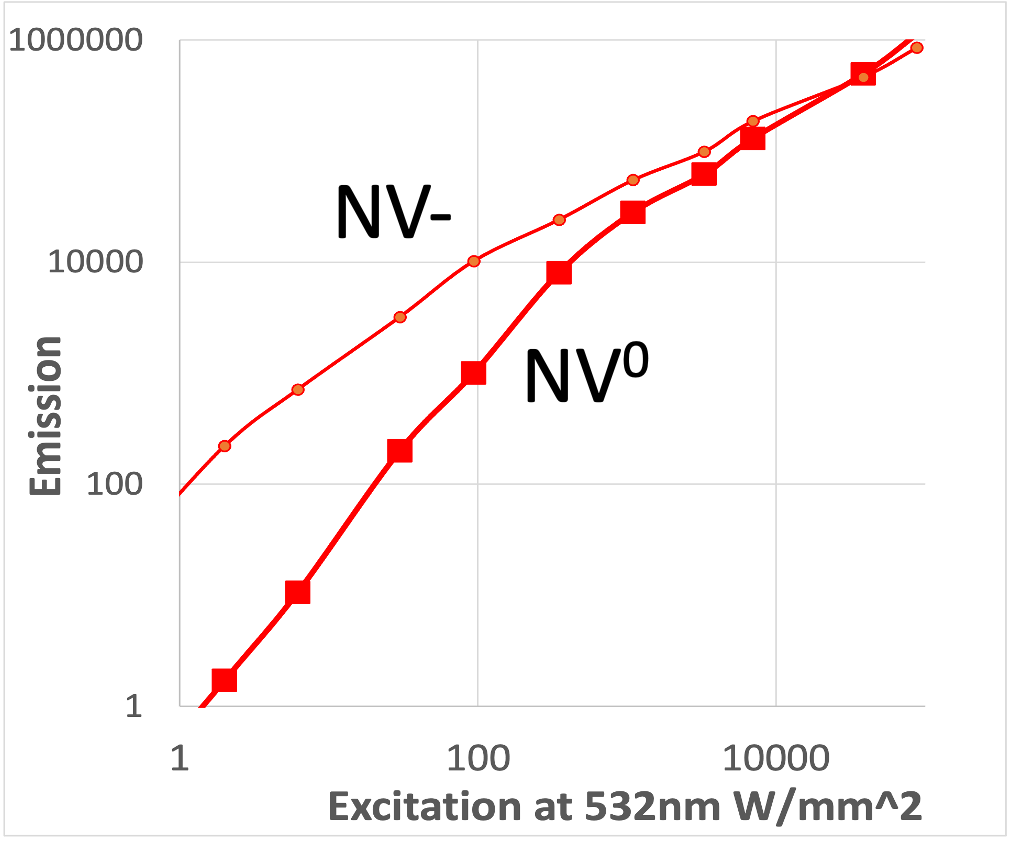} \renewcommand{\thesubfigure}{c}\caption{}\end{subfigure}
\caption*{2. NV${}^0$ created and monitored using 532~nm excitation.}
\caption{\label{exc 600 plus} 1. Various experimental traces to show that the NV$^-$ $\rightarrow$ NV$^0$ charge transfer is linear in excitation intensity. (a) Using the sample with 115 ppm nitrogen there is no NV$^0$ emission at low intensities as in Fig \ref{samples} and (a) here.  (b) With excitation at 620 nm NV$^0$ is created and can be detected using the weak 532 nm probe. Filters eliminate response > 600 nm. (c) This NV$^0$ response is a linear function of the 620 nm excitation intensity shown for 0 to 600 mW. 2. When 532 nm is used to excite NV$^-$ and create NV$^0$ and also to detect the NV$^0$ emission both excitation and detection are linear. The result is that the response will be quadratic, and this is the situation at low intensities 0 -- 0.5~W/mm as shown in (b). There is no obvious change towards higher intensities in (c) but functional form is not conclusive}
\end{figure}

\subsection{NV charge state: Tunneling transition J}
There is a very significant reverse transfer from NV$^-$ to NV$^0$ that can occur with excitation of NV$^-$.  With a laser at wavelengths between the zero-phonon line of NV$^-$ at 637 nm and that of NV$^0$ at 575 nm the only excitation is that of the $^3$A$_2$ -- $^3$E transition of NV$^-$. Excitation at 620 nm is used as an example and with this excitation it is found that the intensity of NV$^0$ created is linearly dependent on the intensity of the excitation laser. The situation regarding charge state is monitored using a weak probe at higher energy of 532 nm that by itself does not create any NV$^0$ (Fig \ref{exc 600 plus}~(1)). The figure, therefore, is consistent with transition J in Fig \ref{pair} being spontaneous.  The situation is also investigated for increasing 532 nm excitation but in this case the excitation and monitoring of NV$^0$ are both linear and a quadratic dependence is observed as anticipated (Fig \ref{exc 600 plus}~(2). As intensity is increased the contribution of transition K becomes significant and the response is close to linear depending on the balance between J and K.)

\begin{figure}[ht!]
  \centering
    \begin{subfigure}[b]{0.596\textwidth}
        \includegraphics[width=\textwidth]{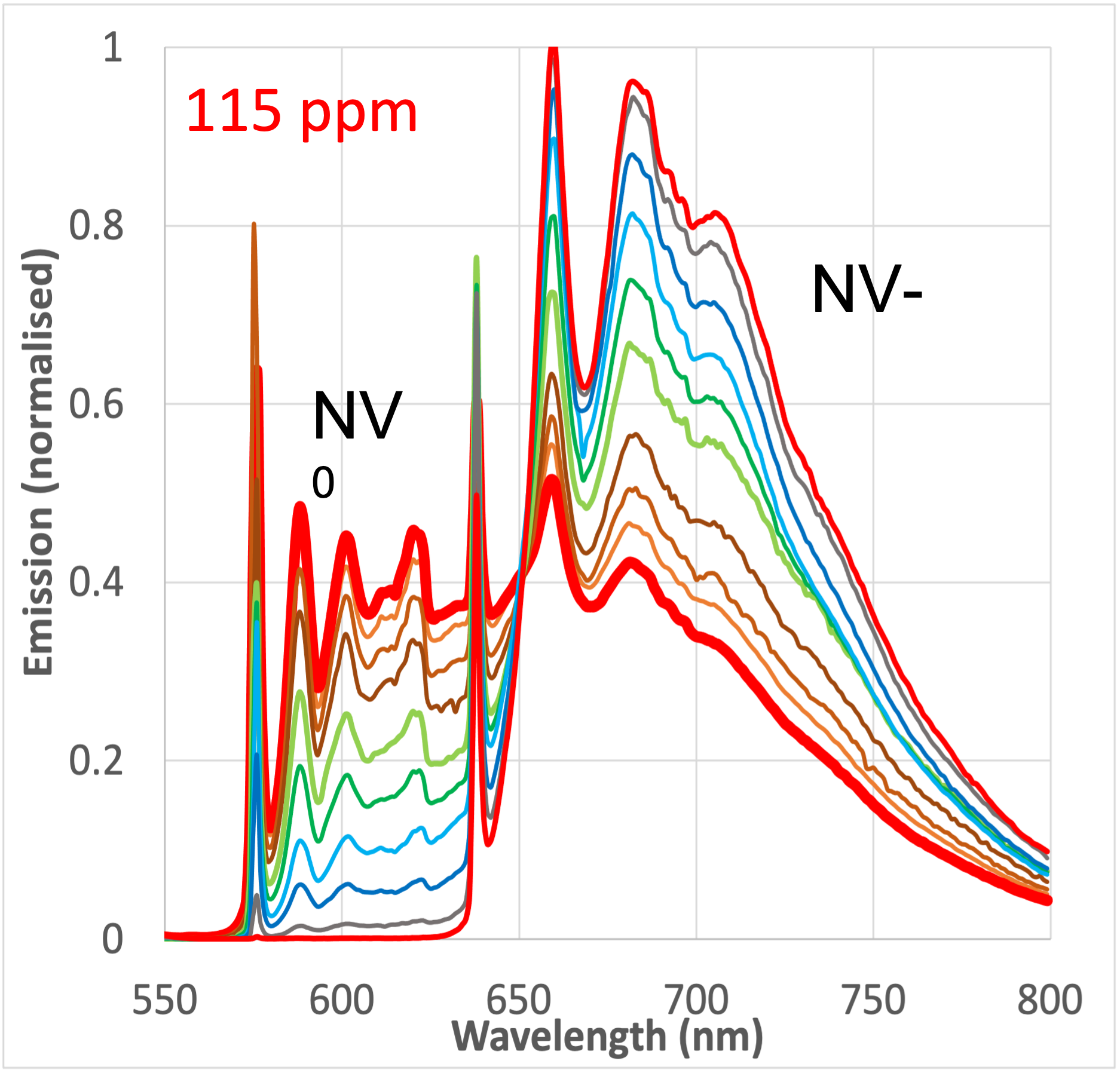}
        \caption{}
        \label{}
    \end{subfigure}
 ~
\begin{minipage}[b]{0.37\textwidth}
\centering   
    \begin{subfigure}[b]{1\textwidth}
        \includegraphics[width=1\textwidth]{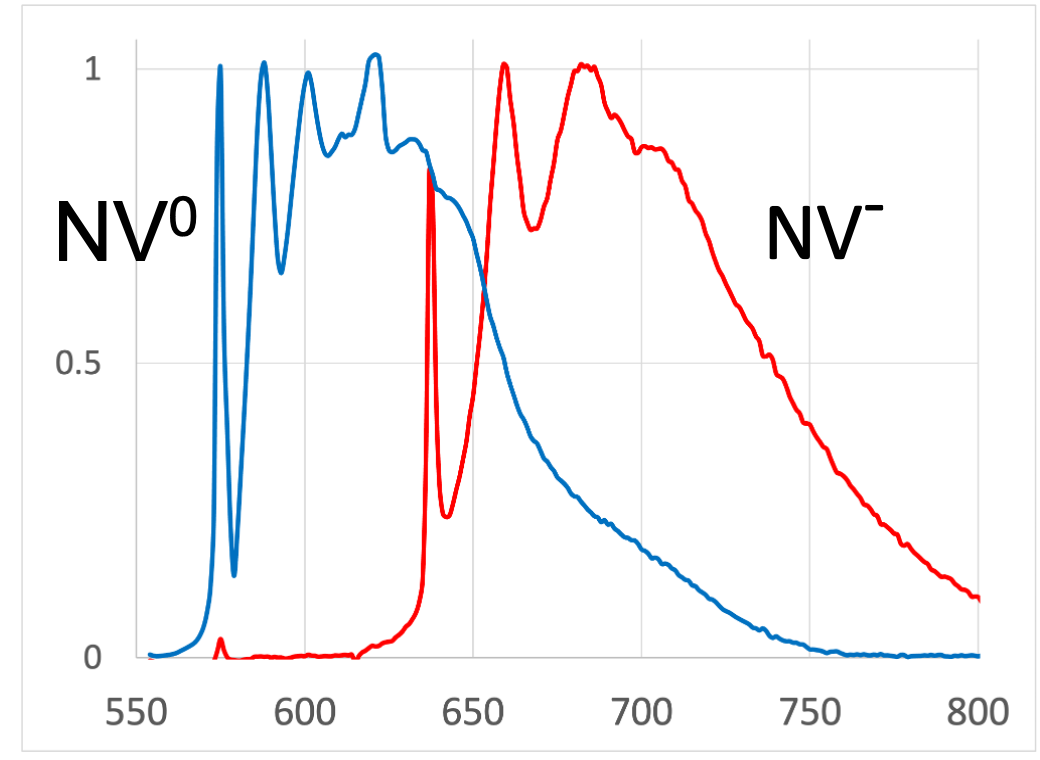}
        \caption{}
        \label{}
    \end{subfigure}

    \begin{subfigure}[b]{1\textwidth}
        \includegraphics[width=1\textwidth]{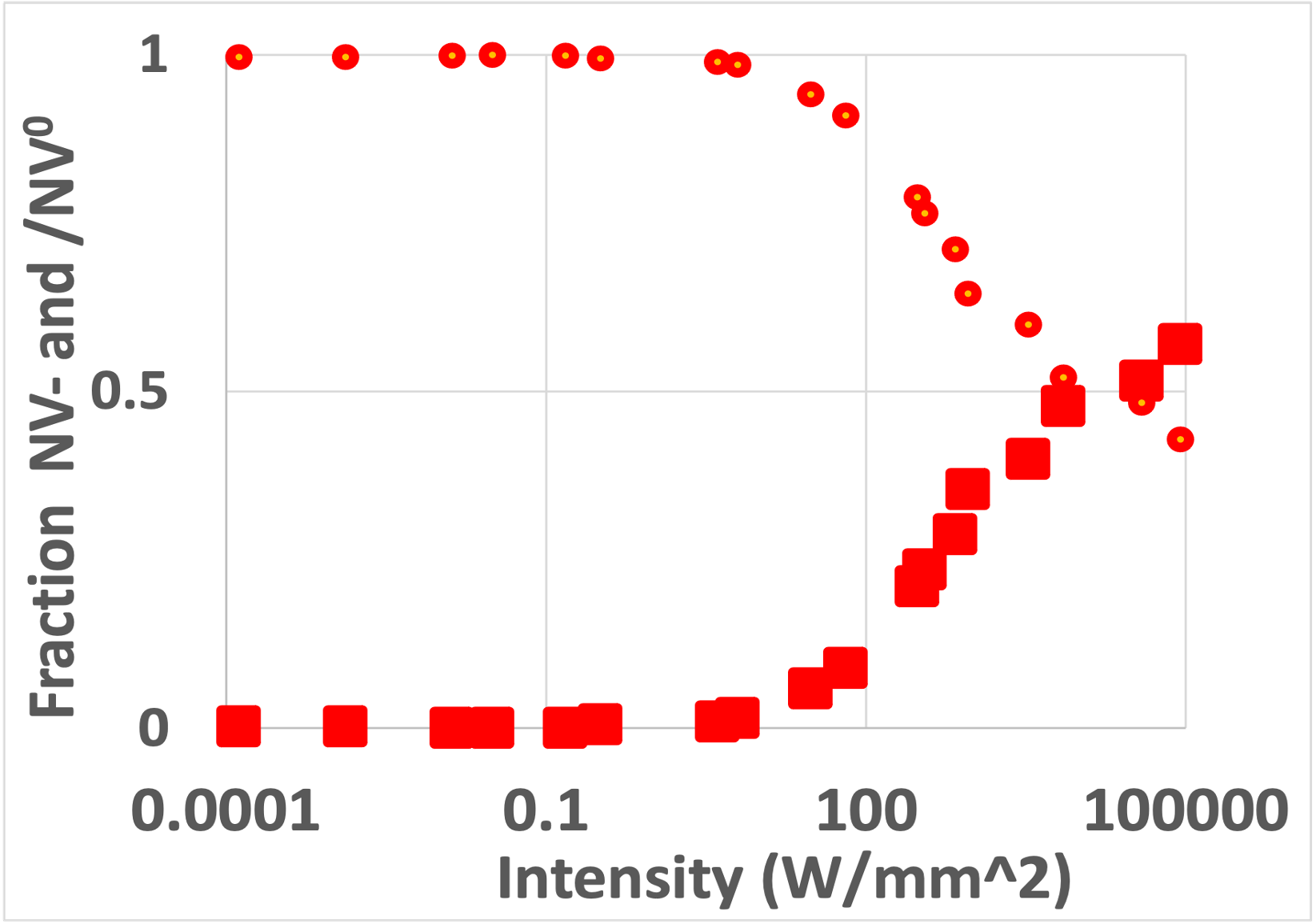}
        \caption{}
        \label{}
    \end{subfigure}
\end{minipage}
\caption{\label{rand}  Figure shows the increase of NV$^0$ with increasing excitation intensity illustrated for the 115 ppm sample. (a) 77 K emission of the 115 ppm  N sample for excitation intensities from 1 mW/mm$^2$ to 10$^5$ W/mm$^2$. The traces are normalise for constant area. The illuminated area varies from 1 mm diameter for lowest intensities, to 100 micron  for intermediate intensities to 3 micron  for highest intensities. (b) indicates normalised spectral distribution of NV$^0$ and NV$^-$ and the contribution is determined for each excitation. (c) gives the fraction of NV$^-$ and NV$^0$ emission in each case and presented for varying excitation intensities. There are two processes giving rise to the changes. Only J at low intensity but additional ionisation of the nitrogen (introduced later) at higher intensities although with almost seamless change with increasing intensity as described later in the text.}
\end{figure}

\begin{figure}[ht!]
  \centering
    \begin{subfigure}[b]{0.47\textwidth}
        \includegraphics[width=\textwidth]{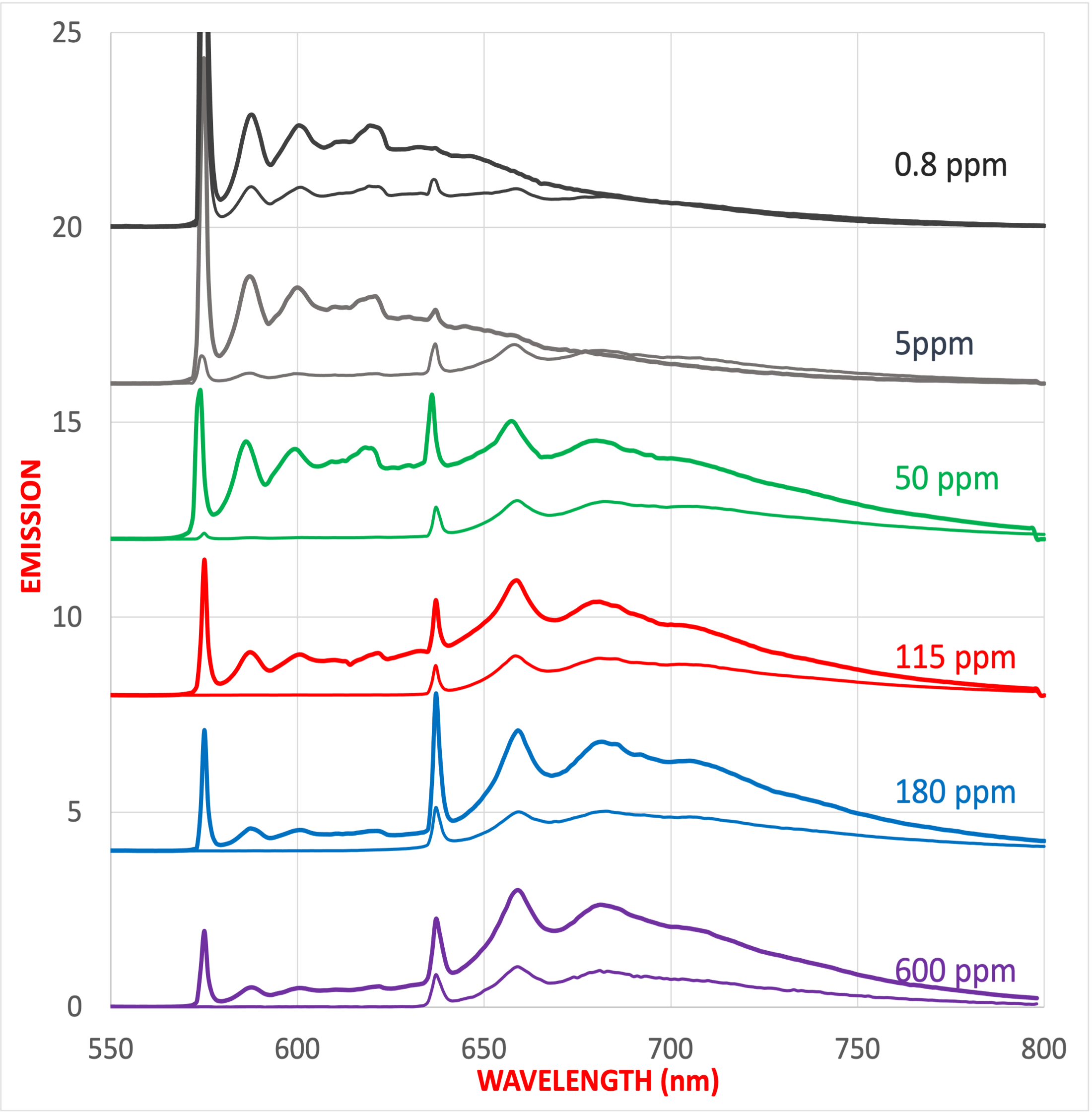}
        \caption{}
        \label{}
    \end{subfigure}
~   
    \begin{subfigure}[b]{0.45\textwidth}
        \includegraphics[width=\textwidth]{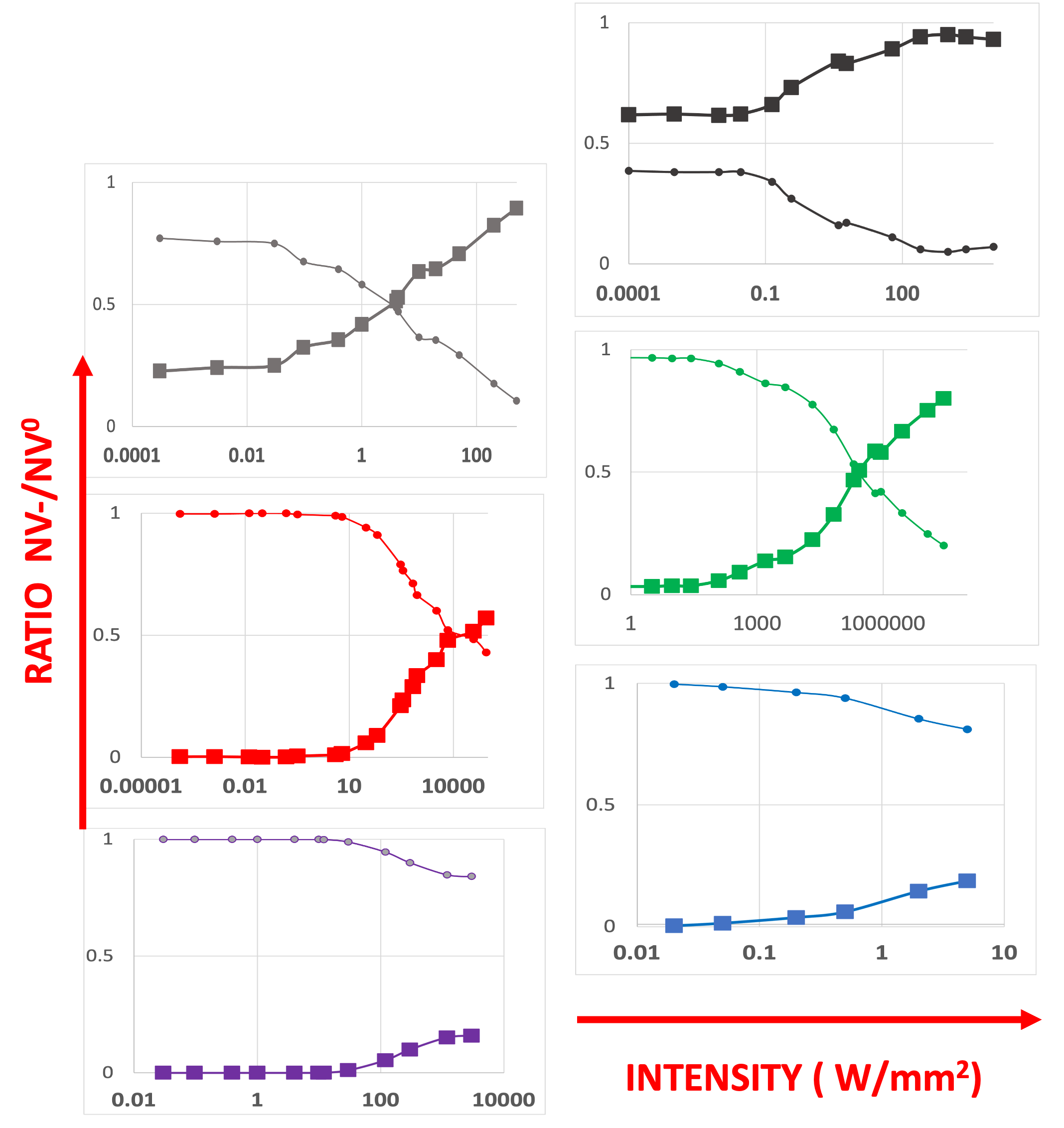}
        \caption{}
        \label{}
    \end{subfigure}
\caption{\label{power} Spectra indicating change of the relative intensities of NV$^0$ and NV$^-$ between low and high intensities for the samples given in Fig \ref{samples}. The top sample in Fig \ref{samples} was damaged and not included. (a) Representative spectra for excitation at low and high intensities (not same intensities for each).  NV$^0$ emission to short wavelengths of 637 nm and NV$^-$ to  longer wavelengths. (b) Indicates the change of the NV$^0$ /NV$^-$ ratio over many orders of magnitude of excitation intensity. All samples show the same trend but clearly with different initial ratio and different rates.}
\end{figure}

\begin{figure}[h!]
  \centering
    \begin{subfigure}[b]{0.22\textwidth}
        \includegraphics[width=\textwidth]{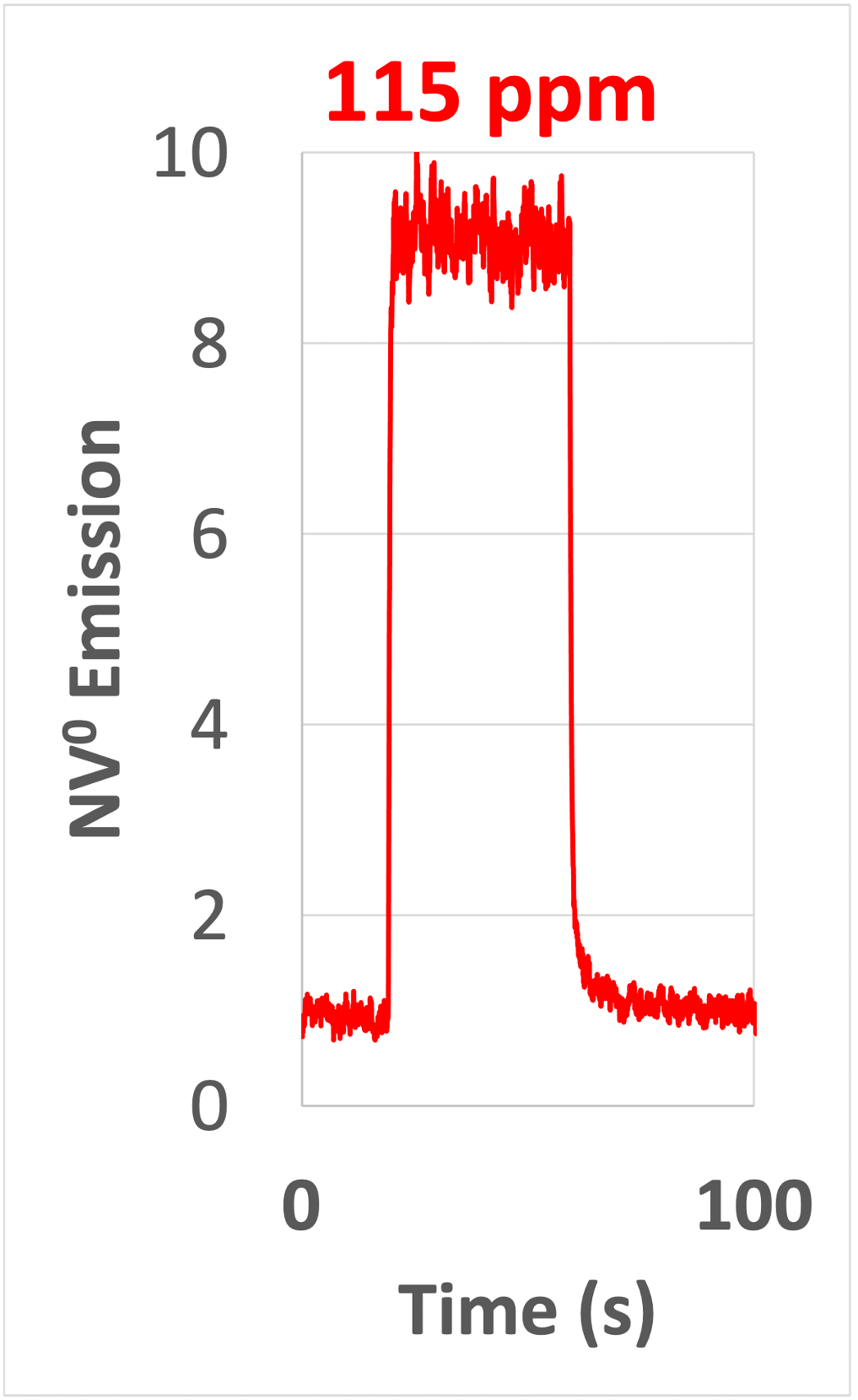}
        \caption{}
        \label{}
    \end{subfigure}
~   
    \begin{subfigure}[b]{0.38\textwidth}
        \includegraphics[width=\textwidth]{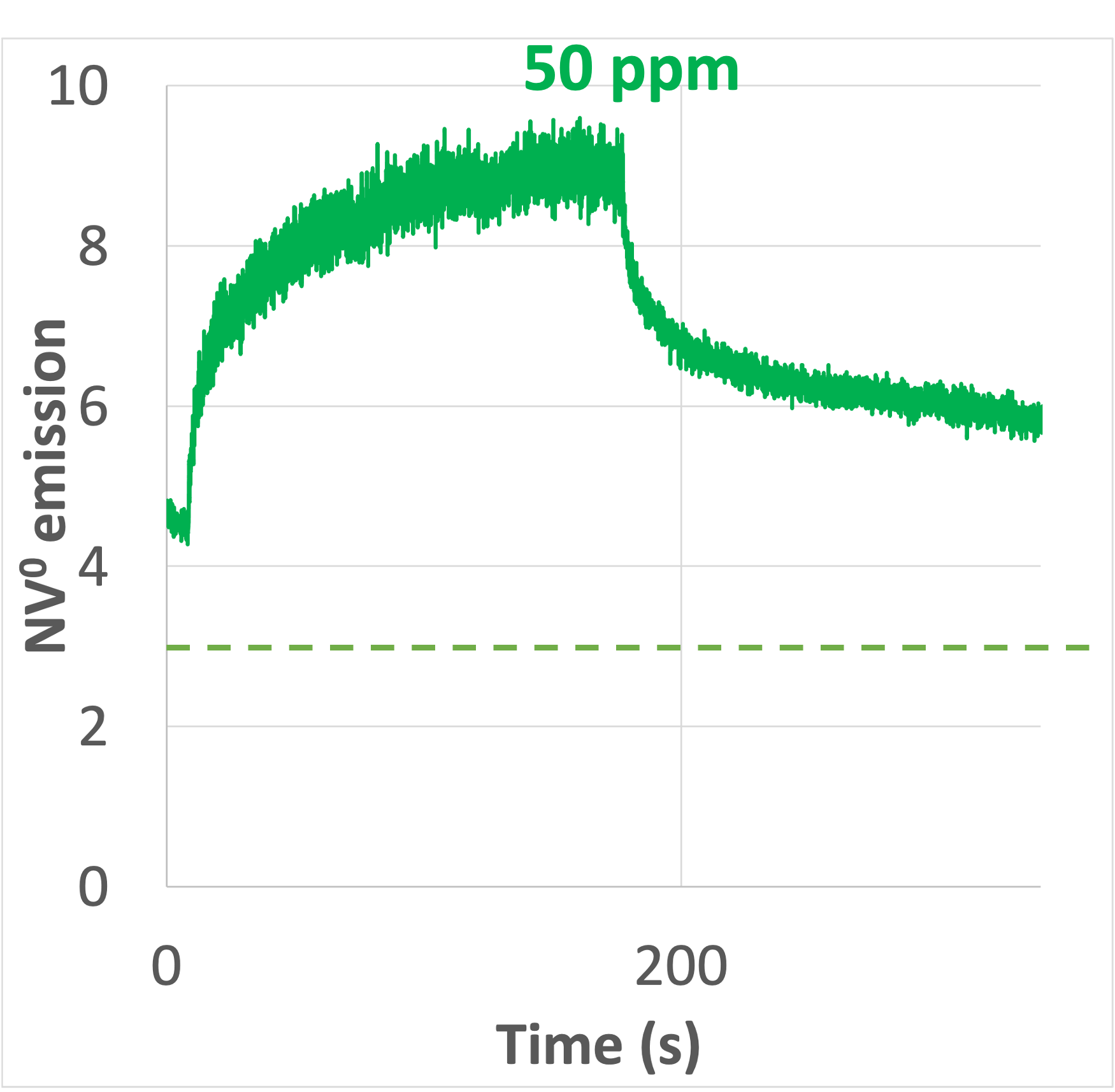}
        \caption{}
        \label{}
    \end{subfigure}

    \begin{subfigure}[b]{0.6\textwidth}
        \includegraphics[width=\textwidth]{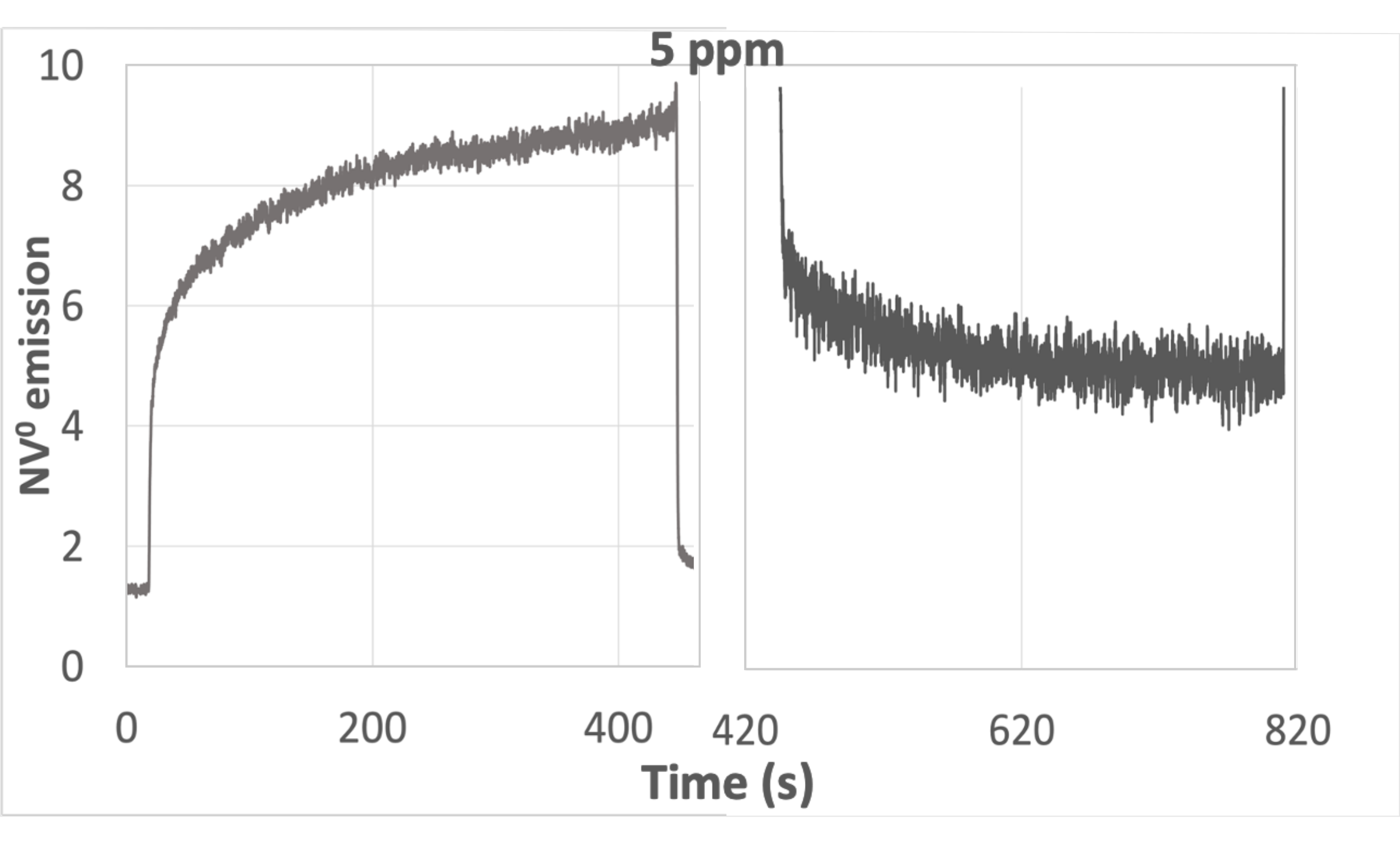}
        \caption{}
        \label{}
    \end{subfigure}
\caption{\label{time}Measurements of NV$^0$ with the excitation of NV$^-$. Traces give NV$^0$ responses for 115 ppm sample, 50 ppm sample and  5 ppm sample created by excitation at 532 nm switched on and off. The presence of NV$^0$ is monitored using a separate very weak 532 nm probe with emission detected at 600 nm. The time scales in the diagrams are similar and the intention is to illustrate the changes of response with nitrogen concentration -- fast for high concentrations (response time restricted by use of mechanical chopping) and slow for low concentrations. In each case with switching on the laser the rate will be largely determined by transition J and when switched off the rate will be set by transition L in Fig \ref{pair}. Similar determination of rates can be obtained by monitoring the NV$^-$ intensities and if taken with a 600 gauss field the rates will indeed be those of the charge transfer as discussed by Gire \textit{et al} \cite{Giri2017,Giri2018}. It can be seen from the traces that they are not simple exponentials and for a given nitrogen concentration there is variation of rates from ms to many seconds. This highlights the issue of attempting to fit spectra with fixed parameters rather than distribution of parameters as mentioned in the Introduction.}
\end{figure}

The NV spectra with increasing excitation intensity are shown in Fig \ref{rand} for the 115 ppm nitrogen concentration sample over a wide range of excitation intensities at 532 nm.  The fraction of NV$^0$ increases with increasing laser excitation (Fig \ref{rand} (c)). At low intensities as described above the process is due to the tunnelling mechanism J. With the higher intensities a second process can also be involves and this will be discussed later.  The situation is the same for all samples as shown in Fig \ref{power}. For the lower nitrogen concentrations towards the top of the figure there is a significant NV$^0$ emission whereas there is only small NV$^0$ fraction for the very high nitrogen concentration towards the bottom of Fig \ref{power}. This trend is, however, misleading as the tunnelling rates will be faster for the higher nitrogen concentrations. The reason for the apparent reversal is that in all cases the NV$^0$ will decay to its ground state and in the ground state the back transfer L to NV$^-$ will again be fastest with high concentrations. Hence the magnitude of the NV$^0$ response in  a CW experiment depends on the rates of both L and J. To illustrate this issue the rate of change of the NV$^0$ emission is shown in Fig \ref{time} for three cases. For the 115 ppm sample the signal with switching on and off laser excitation there is a rapid increase and decreases of the NV$^0$ emission. For even higher concentrations the rates are too fast to record and consequently not shown. However, with low concentration sample 20 ppm and 5 ppm the responses are slow. Once NV$^0$ is created the centres are long lived and indeed the signals are only large because the signal is integrated over the excitation period. 

The transition J in Fig \ref{pair} involving charge transfer from the $^3$E excited state to NV$^0$ shortens the emission lifetime. Also, as the initial transfer is not spin selective any spin polarization in the $^3$E state will be lost with the cycling back to the ground state. Thus  the effect of this charge cycling is to reduces the polarisation as well altering the lifetimes.

\subsection{ NV  charge state: Change due to N ionisation}
\begin{figure}[ht!]
  \centering
    \begin{subfigure}[b]{0.58\textwidth}
        \includegraphics[width=\textwidth]{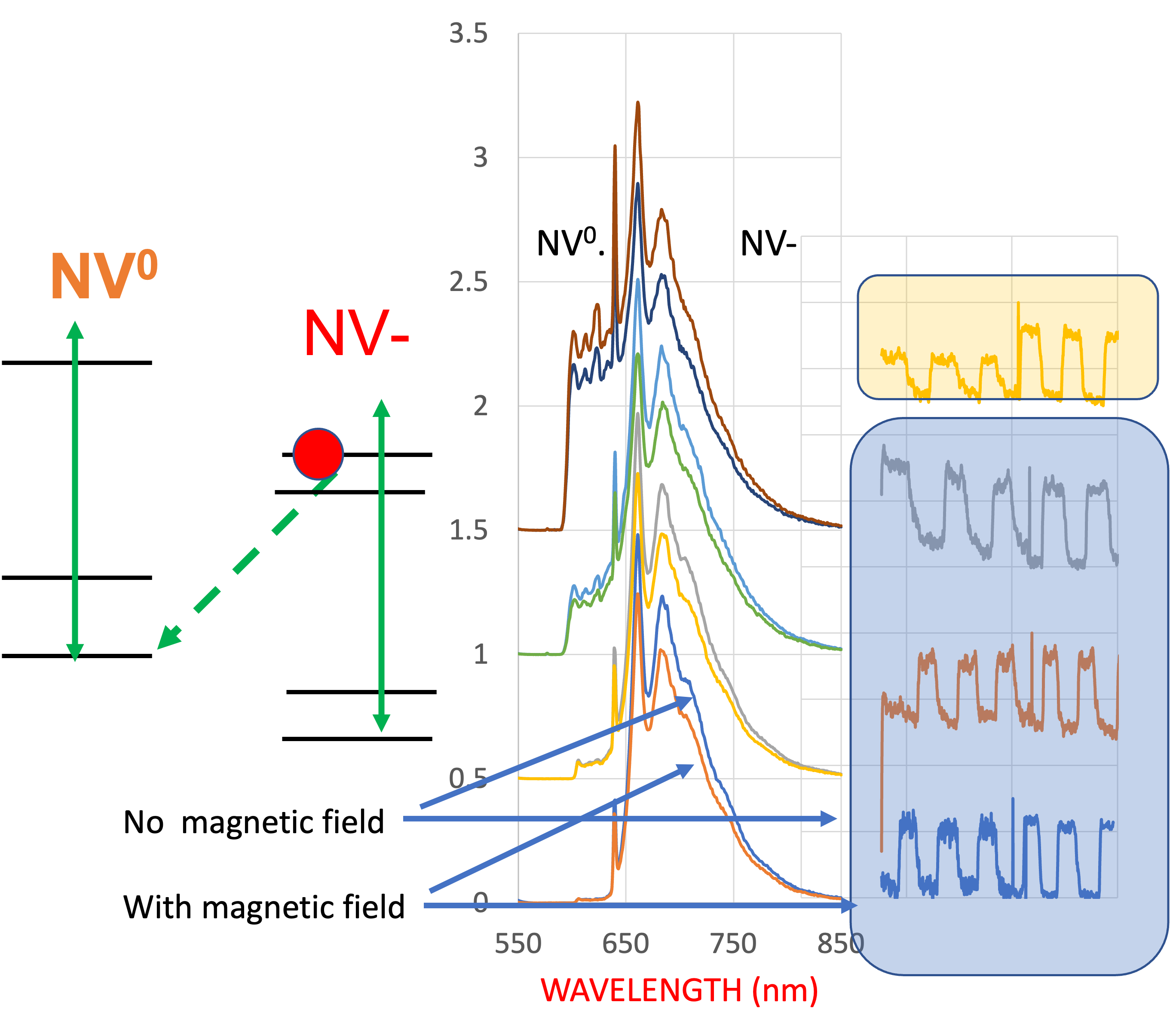}
        \caption{575~nm.}
        \label{}
    \end{subfigure}
~   
    \begin{subfigure}[b]{0.38\textwidth}
        \includegraphics[width=\textwidth]{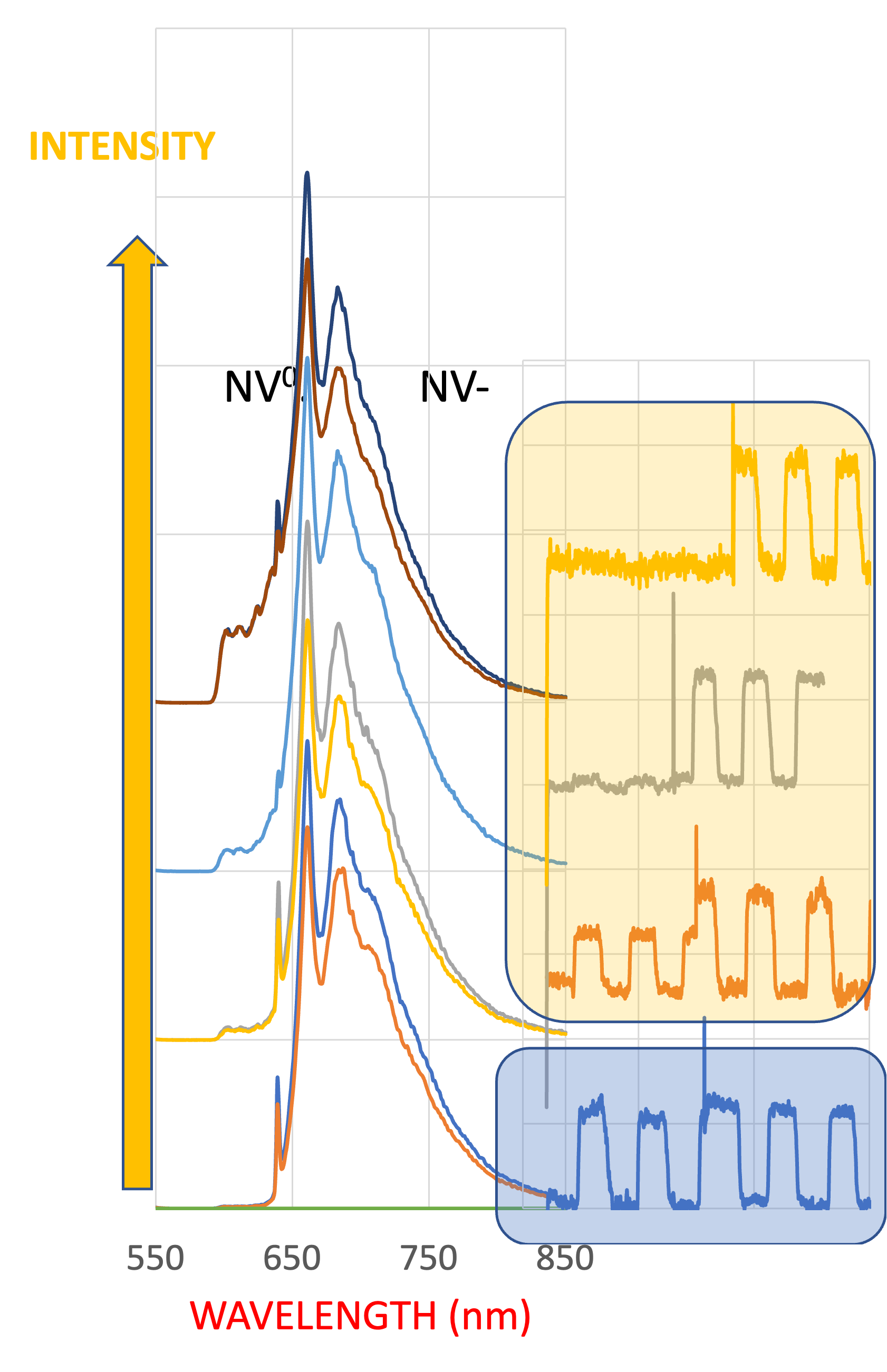}
        \caption{532~nm.}
        \label{}
    \end{subfigure}
\caption{\label{Y and G} Experiments illustrating that the NV$^0$ $\rightarrow$ NV$^0$ charge transfer is linear but characteristics vary with excitation wavelength and intensity. The energy levels on the left indicate that the conversion to NV$^0$ is dependent on the  $^3$E excited state population. This population can be reduced by applying a magnetic field and correspondingly there is a reduction of polarisation and of the NV$^-$ emission. Spectra on the right show the change with the application of a magnetic field (a) 575 nm excitation and (b) 532 nm excitation. Each trace shows a high NV$^-$ response when polarised with no applied field and a reduced response when field is applied. There is also a response for NV$^0$. The variation of the responses is shown for NV$^0$ and NV$^-$, in each case switch on and off magnetic field three times and with matching intensities. The significance is that for 575 nm excitation the NV$^0$ responses match that of NV$^-$ (high-lighted by pale blue) except that for the highest intensities whereas for 532 nm the only one matching is for lowest intensity.  For the matching contrast (light blue) the charge transfer is linearly dependent on the $^3$E population and the charge transfer is due to the spontaneous transition J in Fig \ref{pair}. The other situation high-lighted in yellow involve ionization of the nitrogen and the mechanism is discussed in the text.}
\end{figure}

With increasing intensity there is a second mechanism that affects the charge transfer, and the mechanism varies with wavelength as well as intensity. The situation is best studied and illustrated using magnetic fields and examples are shown in Fig \ref{Y and G}.

An off axis magnetic field mixes the spin states and effectively quenches spin polarisation. Magnetic field studies are treated more fully later but here the significance is the application of the field quenches spin polarisation and reduces the excited state population (left diagram in Fig \ref{Y and G}). The emission intensity originating from the excited state is reduced (spin contrast). It follows that for the mechanism J where the decay to NV$^0$ is also from the excited state the NV$^0$ created will be reduced by the same amount. This is what is observed for the spectra using 575 nm excitation Fig \ref{Y and G}(a) and for very  low intensities for 532 nm in Fig \ref{Y and G}(b). The contrast of NV$^0$ and NV$^-$  are highlighted in light blue in Fig \ref{Y and G} and the contrast of NV$^0$ and NV$^-$ are the same. The exceptions are at the highest intensity using 575 nm excitation and for almost all intensities using 532 nm excitation and in these cases the NV$^0$ contrast is reduced when the magnetic field is applied.  For even modest excitation of 532 nm the contrast is zero implying there is no change of the NV$^0$ emission with magnetic field. The effect is due to the second charge transfer mechanism described below.

The extra charge transfer mechanism is attributed to the ionisation of the bath nitrogen. Single substitutional nitrogen lies 1.7 eV below the conduction band and so there can be ionised with energies > 1.7 eV although require energies > 2.2 eV to be significant (Fig 1 in reference \cite{Rosa1999} and \cite{Nesladek1998}).  Thus with 532 nm (2.32 eV) the ionisation of the nitrogen bath is pronounced. With 575 nm (2.15 eV) the ionisation is much less and only obtained here at high intensities. The mechanism for the charge transfer with the ionisation of the nitrogen is similar as to what occurs with UV excitation.  It is generally known that for defects in diamond with nitrogen as donors that UV excitation creates both donor and nitrogen in the neutral charge state and, hence, in the NV case gives NV$^0$ and N$^0$. This has been shown in references \cite{Lu2017} and \cite{Li2022}. In both cases samples investigated only gives NV$^-$ emission with visible excitation but when exposed to UV there is clearly a conversion to give NV$^0$. The charge transfer is attributed to ionisation and subsequent relaxation of the nitrogen. There is no excitation of NV$^-$ and no effect on lifetime from this process. The N$^+$ created will relax involving the predominant substitutional nitrogen \cite{Ulbricht2011} but more significantly can also alter  the charge of the NV$^-$ impurities.

\begin{figure}[ht!]
	\centering
	\includegraphics[width=1.0\textwidth]{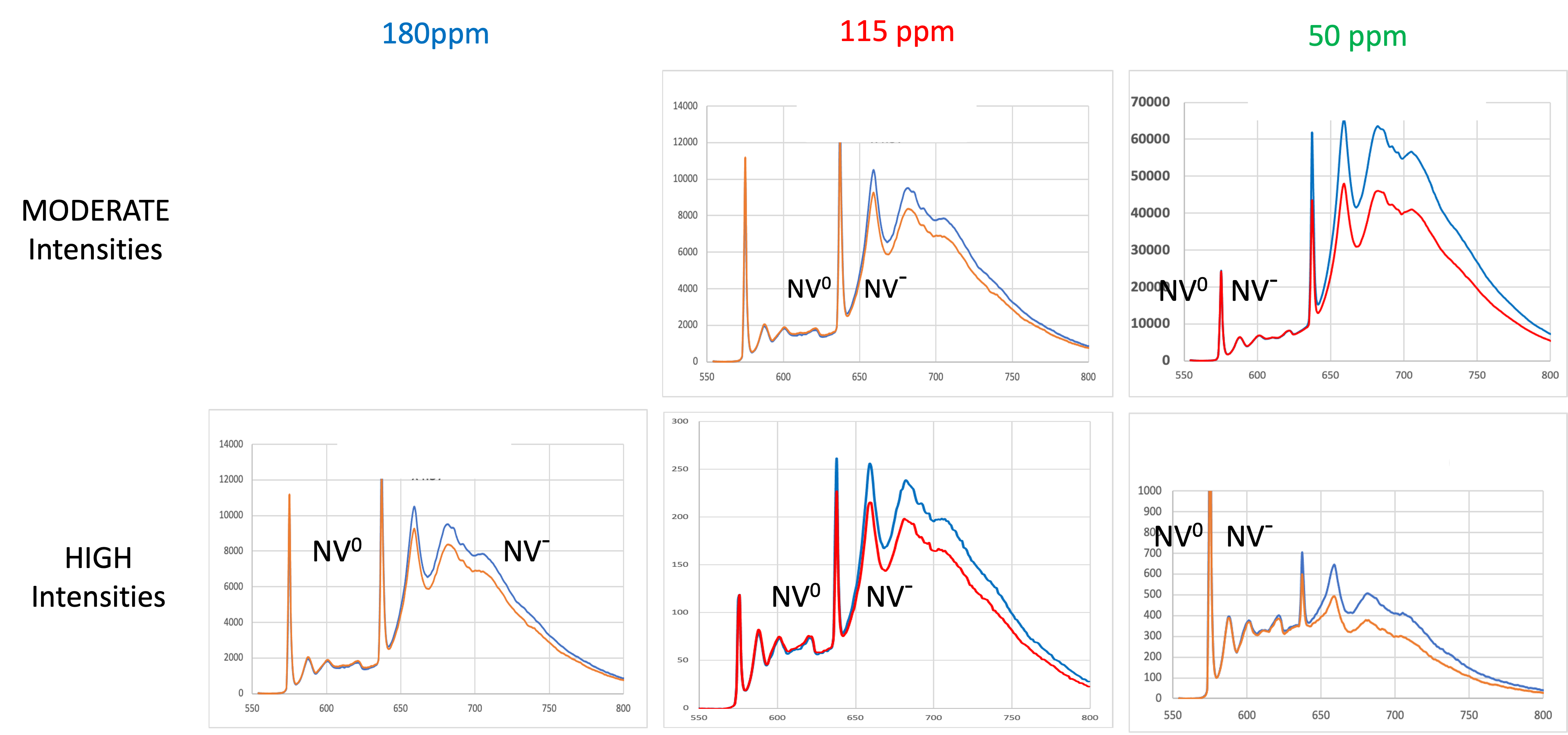}	
	\caption{\label{NV0 samples} Traces give the 77K emission spectra ( similar to Fig \ref{power} ) of various samples with (red) and without (blue) magnetic field (600 gauss along <001> and for two intensities -- `medium' (upper $\approx10$~KW/mm$^2$) and `high' (lower $\approx1000$~KW/mm$^2$). The excitation is at 532 nm and in all cases the application of  a field does not change the intensity of NV$^0$ implying that any charge transfer to NV$^0$ is a consequence of the ionisation of the nitrogen bath.}
\end{figure}

The difference with a magnetic field observed here is useful as it provides a straightforward means of determining the dominant mechanism; Matching NV$^0$/NV$-$ contrast indicates transition J whereas no contrast indicates the second mechanism of nitrogen ionisation. In this way it is found that in the measurements earlier in Fig \ref{power} where 532 nm have been used the ionisation of the nitrogen ionisation is nearly always the dominant origin of change of charge state. Several examples of this situation showing the creation of NV$^0$ with and without magnetic field for different samples and various intensities are shown in Fig \ref{NV0 samples}. In each case the field does not change the NV$^0$ intensity.

Another factor that can be deduced from the magnetic field measurements associated with transition J and observation of the matching contrast of NV$^0$ of NV$^-$. This is that the decay is dependent on the total excited state population and indicates that the  charge  transfer  is independent  of  spin projection. Accepted that there will be a larger transfer from m${}_s =0$ states than from m${}_s=\pm1$ states but only because with spin polarisation the m${}_s=0$ population is larger and not because of different spin-dependent rates.

\section{ Spin polarisation and contrast:  Optical cycle}

This section first summarises the energy level scheme and defines the transitions for the NV$^-$ useful for interpreting experimental data. This is followed by measurements of six samples of various nitrogen concentrations (Fig \ref{samples}) that include contrast for the optical, infra-red and the relative infrared/visible emission ratios. Measurements of lifetimes are also given and  compared with single site lifetimes and of related materials. Hence there is a large amount of data reporting the trends with nitrogen concentration. However, later there is a Summary section that give a more succinct presentation highlighting how the properties vary from low to high nitrogen concentrations. 

\begin{figure}[ht!]
	\centering{\large }
	\includegraphics[width=1.0\textwidth]{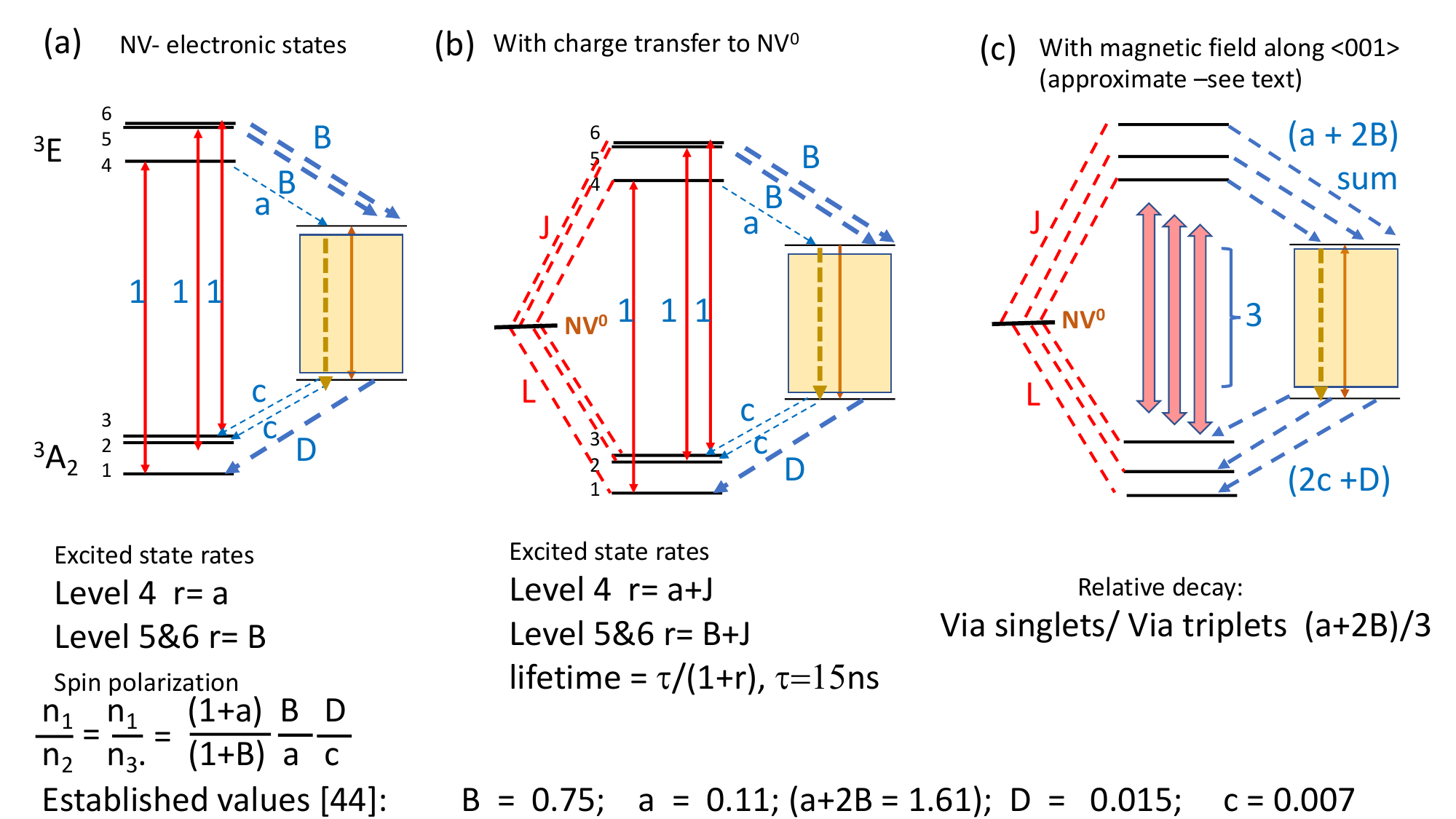}
	\caption{\label{centres} NV$^-$ optical cycle; spin polarisation and normalised transition rates.  (a) Gives the transitions and intersystem crossing rates. Rather than the conventional  k$_i$$_j$ intersystem parameters a, B, c, D (normalised relative to the optical transition as unity) are used indicating that the treatment is not always rigorous and the less formal notation is appropriate. Capitals are used where the parameters favours transfer of population to the m$_s$ = 0 spin projection. The resultant spin polarisation with CW excitation is given below the figure and applies to single site or ensemble just with different magnitude of the parameters (see Table 1). (b) Includes the case where there is charge transfer to NV$^0$ at a rate J and relaxation back to NV$^-$ at rate L. Spin polarisation would include all parameters but no analytical expressions are given in this case. (c) Indicates the situation when a magnetic field is applied along the <001> crystallographic direction. Relationships for the infrared and visible transition rates and the optical lifetime normalised to the strength of the optical transition are given below the figure. Optical transition rate is 66 MHz corresponding to a lifetime of $\tau = 15$ ns.  Below the figure rates for single NV$^-$ centre taken from the reference \cite{Tetienne2012} given in the literature.  Other single centre values reported are given in Table 1.}
\end{figure}

For NV centres it is the intersystem crossing that sets both the lifetimes and the spin polarisation.  The basic optical cycle is same for single and ensemble and a summary diagram is given in Fig \ref{centres}(a). The energy level scheme is the same as NV$^-$ in Fig \ref{single} and \ref{pair} but with the singlet levels displaced to right. From the m$_s$=$\pm$1 excited state the decay to the $^1$A$_1$ singlet is allowed by spin-orbit interaction \cite{Goldman2015a,Goldman2015b} and the relaxation rate is consequently faster than that for m$_s$=0. The difference in rate is significant as it results in (i) preferential population of m$_s$=0 (spin polarisation) and with CW excitation the relative populations are given below the figure.  The rate variation also means  (ii) the emission intensity from the m$_s$= 0 spin state is larger than  that from m$_s$=$\pm$1 and the difference enables the read-out of the spin state.

Fig \ref{centres} (b) gives the situation adding the charge transfer to NV$^0$ as J and relaxation back to NV$^-$ again with L as given in Fig \ref{pair}. In the experimental studies a magnetic field along a <001> direction is frequently adopted and Fig \ref{centres} (c) indicates energy scheme for this situation. Such a field makes equal off-axis angle to the four crystallographic orientations of the NV centres and causes a mixing of the basis states resulting in reduced emission \cite{Lai2009}. A formal treatment for the application of a magnetic field is given by Tetienne \textit{et al.} \cite{Tetienne2017} and it is shown that for a field 600 gauss there is little resultant spin polarisation \cite{Lai2009}. The quenching of the spin polarisation is the main reason for adopting the magnetic field and enables a simple measurement (field/ no field) of the optical contrast. The field is used in preference to resonant ground state microwaves as the field is constant over the whole sample and quenches spin polarisation in ground and excited state. Microwaves only reduce polarisation in the ground state and with one microwave field only for one $\text{m}_s = \pm1 \leftrightarrow \text{m}_s = 0$ spin transition. Also, with microwaves it is problematic to saturate over the excited volume. Thus, for establishing fundamental processes magnetic fields are simpler and satisfactory for investigating the fundamentals of spin polarisation. It is necessary to have Zeeman splittings as large or larger that fine structure splittings, D$_e$ and D$_g$. However, for fields of 600 gauss along <100> the spin polarisation is negligible being < 5 percent of that in the absence of the field \cite{Giri2017,Giri2018} and \cite{Tetienne2017} and described here as `quenched' or `unpolarised'. It is also desirable that the optical cycling that creates the spin polarisation is faster than the spin lattice relaxation time so that the CW spectra corresponds to a `polarised situation'. This is readily obtained with the temperature of 77 K adopted. Spectra with and without a magnetic field are measured and it will be shown in the next section that the radiative contrast visible and infrared reduces with nitrogen concentration.
		
The application of off-axis magnetic fields to quench spin polarisation have often been used in the above way. Chapman and Plakhotnik \cite{Chapman2013} for example adopted magnetic field rather than microwaves in the use of NV$^-$ for study of biological materials. Singam \cite{Singam2016} has likewise adopted such an approach and has given a detailed comparison between responses involving microwaves and magnetic fields. In the case of bio-materials there is a clear advantage of using magnetic fields as it avoids application of intense microwave radiation. The present authors have previously used magnetic fields in the study of NV to obtain spin contrast between polarised and unpolarised situations \cite{Manson2018}. Also Giri \textit{et al}. \cite{Giri2017,Giri2018} have adopted magnetic fields in a study of nano-diamonds and  bulk to differentiate  between  spin polarisation and charge transfer effects. These authors have presented an excellent description and analysis of the technique. Their nano-diamonds had significant nitrogen concentrations and exhibited characteristics not unlike the bulk measurements presented here and their bulk ensembles correspond to the low nitrogen concentrations. The approach adopted here is then not new but rather a proven successful technique for giving insight into charge transfer and spin polarisation.

\begin{figure}[ht!]
  \centering
    \begin{subfigure}[b]{0.45\textwidth}
        \includegraphics[width=\textwidth]{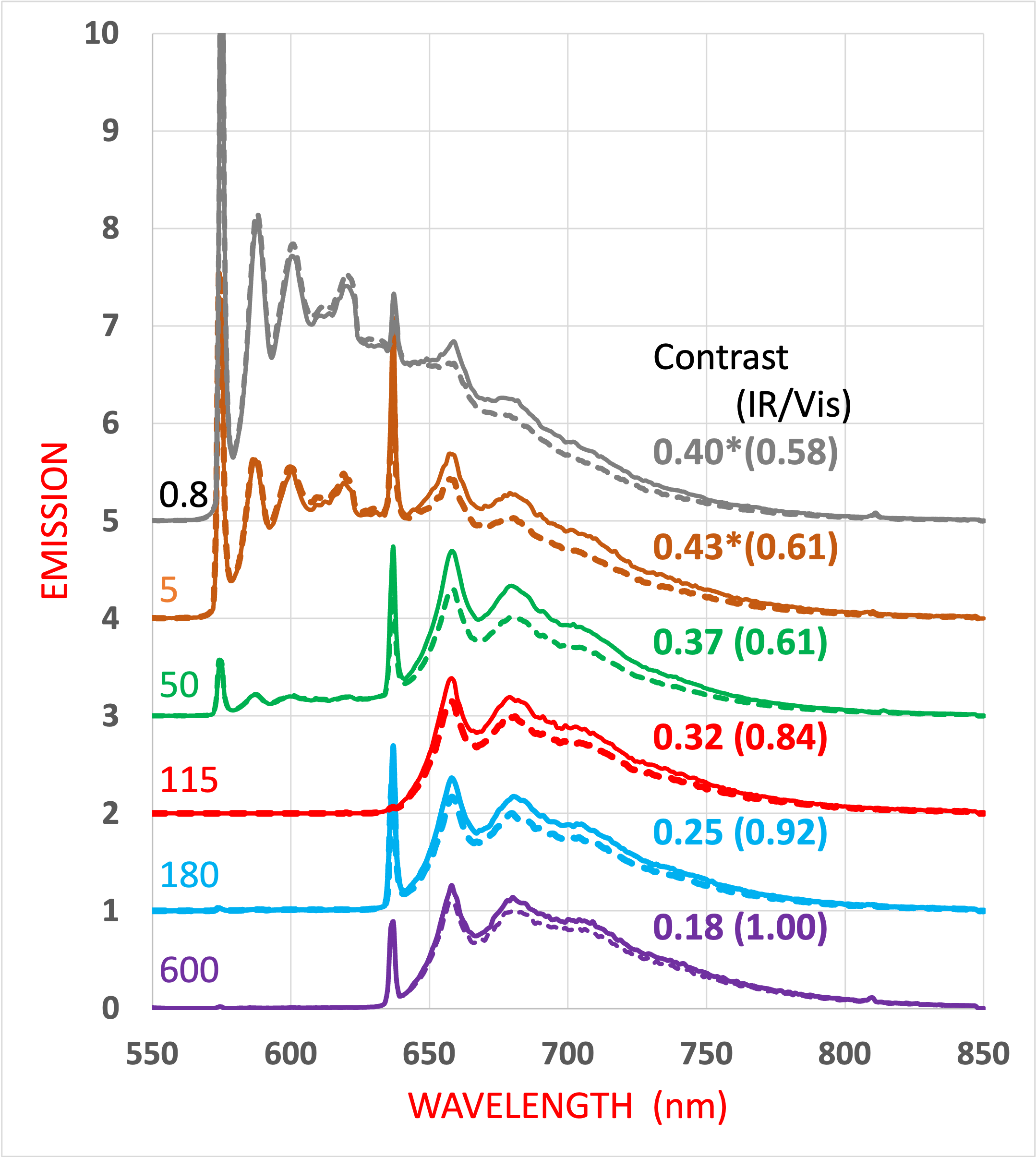}
        \caption{}
        \label{}
    \end{subfigure}
~   
    \begin{subfigure}[b]{0.457\textwidth}
        \includegraphics[width=\textwidth]{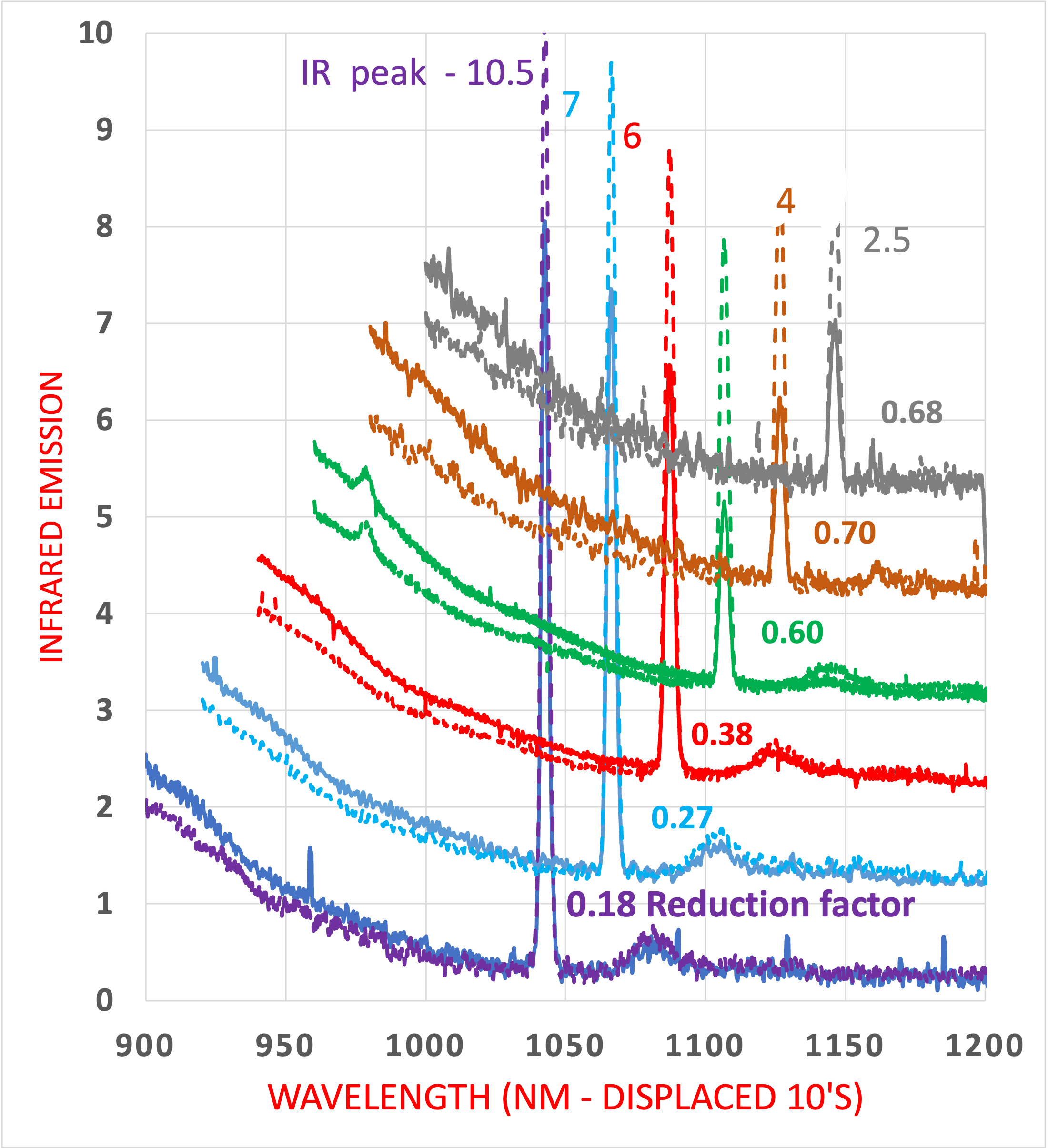}
        \caption{}
        \label{}
    \end{subfigure}
\caption{\label{field ir} The emission spectra of six samples with (dashed) and without (solid line) a 600 gauss magnetic field applied. The visible spectra are on the left (a) and infrared on the right (b). The infrared traces are normalised to that of the extreme long wavelength component of the $^3$E -- $^3$A$_2$ emission at 950 nm. The numerical values associated with the infrared traces in (b) give the contrast as the fractional reduction of the emission with polarisation. The values for the visible give the equivalent increase of emission. There is a correspondence between these responses (see text) and division of the (a) value by (b) value gives the total decay via the singlets relative to that via visible. For example, for the 115 ppm sample, in red the emission increase by a fraction of 0.32 with the polarisation whereas for the infrared the $^1$A$_1$ -- $^1$E zero-phonon line at 1042 nm decreases by 0.38.  Dividing the visible increase 0.2 by the infrared reduction of 0.35 gives total infrared decay relative to the visible decay and this value 0.84 is included in brackets in (a). As described in the text it is this value that corresponds to the ratio of (a+2B)/3. The asterisk in the top two traces indicates that the contribution from the NV$^0$ emission has been subtracted but there still maybe some error.  For convenience other data included in the figure is the peak infrared emission (also unpolarised) More detailed data is given later in Fig \ref{fourspectra} but for a smaller number of samples.}
\end{figure}

\subsubsection{Spin polarisation and contrast:  Measurements using  magnetic field }
Many of the significant experimental observations are included in Fig \ref{field ir} (a) and (b). These include contrast visible and infrared with and without applied magnetic field. There a decreasing contrast with increasing nitrogen and values are given on the spectra traces. What is also clear from the traces is that the infrared emission increases relative to the visible with increasing nitrogen (also see Fig 28(b) in reference \cite{Manson2018}) and from Fig \ref{centres}(c)  this implies a change of the intersystem crossing. An attempt to obtain a quantitative measure of the change from the fraction decay via the singlets compared to that via the $^3$E -- $^3$A$_2$ radiative transition is given in the next paragraph.
		
The spectra for Fig \ref{field ir} are taken for weak excitation (10$^{-5}$ of saturation factor) such that there is only small changes to the  population in the ground state and only minor population in the singlet levels. There is no change to this situation with the application of the magnetic field. The laser excitation is the same with and without magnetic field and hence the population transferred to the excited state will not be altered  by the magnetic field. The excitation is constant and with no change of population in other states it follows the total decay for field and no-field will be constant. The increase of the visible emission will be matched by the decrease of decay via the singlets.  It is recognised that the infrared emission is just a small part of the total decay via the singlets (`the peak of the iceberg') but the fraction, radiative to non-radiative, will not be altered by the magnetic field. The change of the infrared emission indicates the change of decay via the singlets just with a scale factor. The fractional change from the unpolarised responses is given adjacent to each trace -- increase in the case of the visible and decrease for the infrared.  Knowing the fraction of the infrared decay in relation to visible enables the total singlet decay to be determined and this is the value given in brackets in Fig \ref{field ir}(a) for each sample. From the transitions indicated in (i.e. unpolarised) Fig \ref{centres} (b) there will be equal population in the three excited states and the total decay via the singlets has a value of (a+2B) relative to 3 for the visible $^3$E $\rightarrow$ $^3$A$_2$ emission Fig \ref{centres} (c). The ratio, therefore, gives a measure of the sum of parameters (a+2B)/3 and the experimental values are given in brackets in the last column in Table 1. These values are compared with (a+2B)/3 values calculated from the lifetimes reported in the next section.
	
There is overall consistency in the trends of the change of the upper inter-system crossing but the detailed values (a+2B)/3 obtained using the two different approaches are not in agreement. For example, values calculated from lifetimes are of order of 1.5 higher than the above experimental approach. Also, the change with nitrogen concentration from low (5 pm) to high (181 pm) varies by over a factor 2.1 using lifetimes and only 1.55 from experiment. The difference in the two approaches suggest that there may be a non-radiative decay not observed with optical detection. However, there is concern with the very different techniques and the measurements involving changes of the optics, illumination of area on samples and detectors. Clearly there needs to be further investigation of this difference.
	
\begin{figure}[ht!]
\centering
\includegraphics[width=1.0\textwidth]{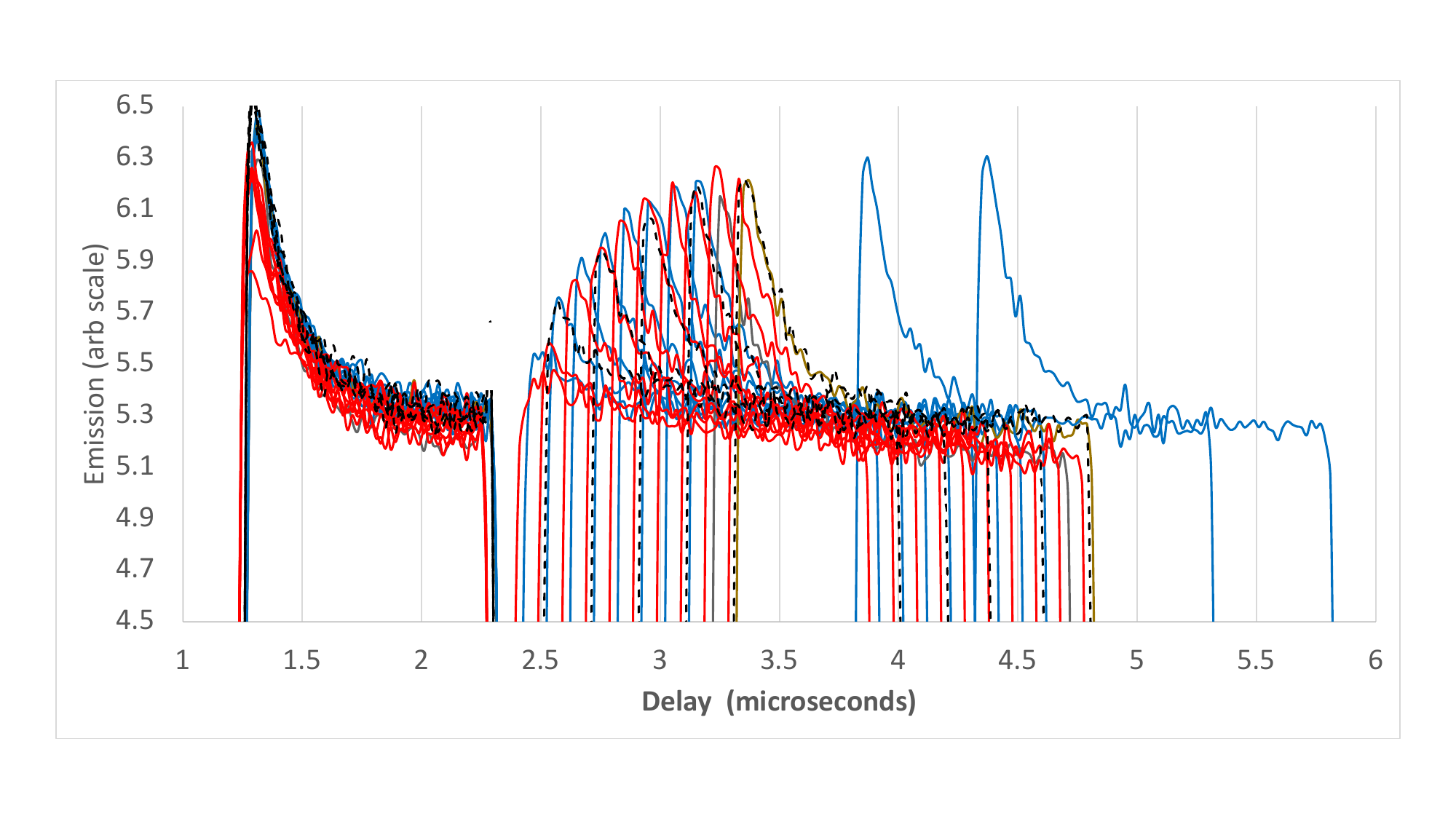}
\caption{\label{singlet E}  Experiments that indicate that there is no significant change to the lower  singlet-triplet intersystem crossing with nitrogen concentration. The traces are for three samples with varying nitrogen concentrations Blue corresponds to 180 ppm nitrogen sample, red to 20 ppm sample and the dashed trace is of an intermediate concentration of 115 ppm. It is appreciated that the traces overlap but this clearly indicates that there are no differences in the lifetime of the lower singlet level with nitrogen concentration.}
\end{figure}
	
All of the above involve studies are  of the upper inter-system crossing. Possible changes of the lower intersystem crossing were also considered and for this the singlet level $^1$E lifetime was measured. The measurement involves the recovery of emission from the case where population is transferred and stored momentarily in the $^1$E state. Such measurements have been reported by Manson \cite{Manson2006} and Robledo\cite{Robledo2011}. The $^1$E lifetime (although temperature dependent) was found to be independent of the nitrogen concentration and at 77K the rate of $\approx$ 500 ns corresponds to that given by Robledo\cite{Robledo2011}.  It is concluded that (2c+D) is not changed with nitrogen concentration.

The conclusion is that whereas there is no change to  the lower crossing (2c +D) there is a clear change of  the upper intersystem crossing (a+2B) by near a factor of 2 between low to high nitrogen concentrations.  There has not been a determination whether a or B or both change. Further information on the changes to a and B can be determined form the lifetimes and this is the focus of the next section.

\subsubsection{ Spin polarisation and contrast: Change of optical cycle}
\begin{figure}[ht!]
\centering
\includegraphics[width=1.0\textwidth]{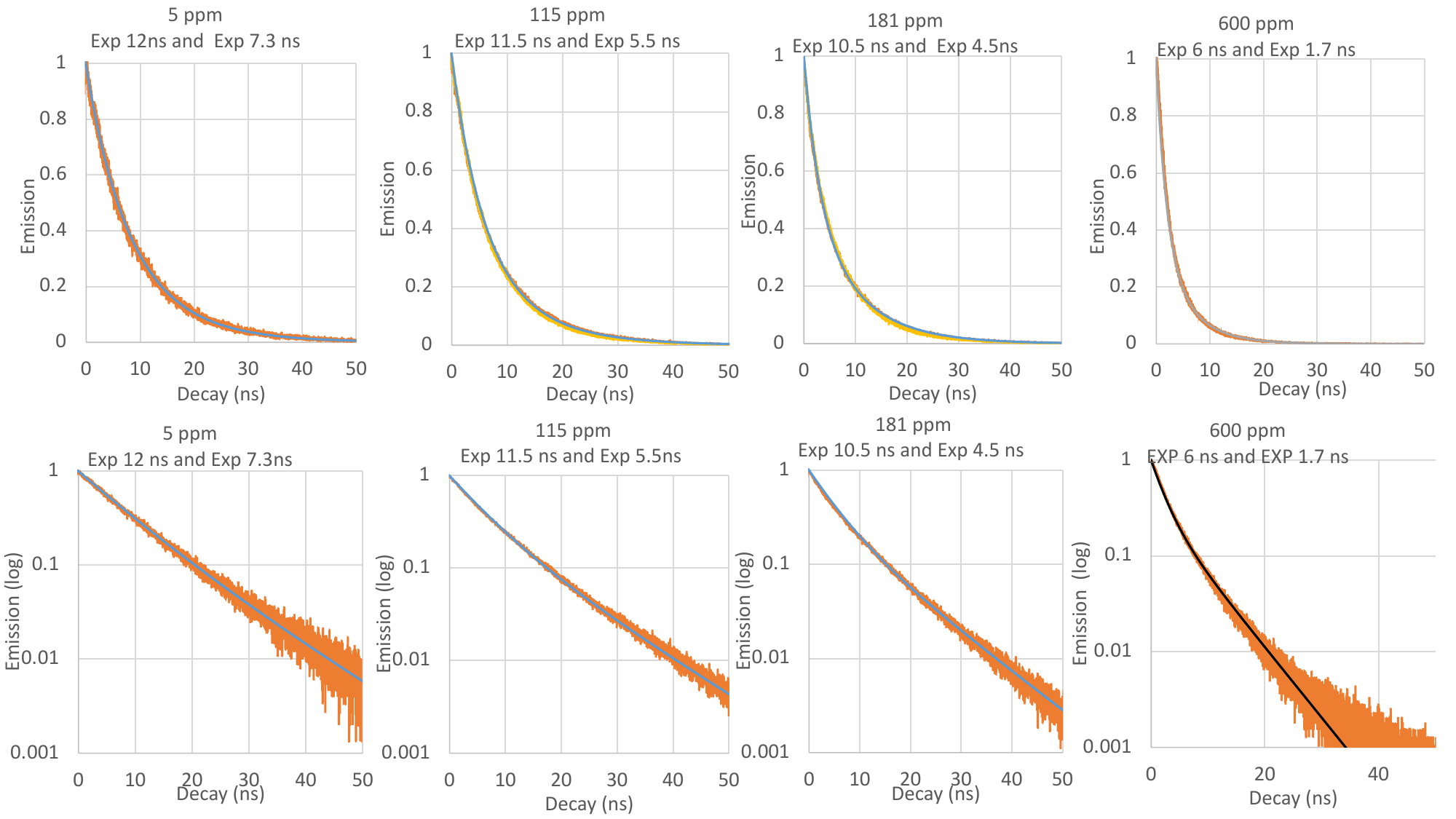}		
\caption{\label{lifetimes} Experimental observation of room temperature lifetimes and fit to dual exponential decays. The observed emission lifetime are given in upper traces and log plots in lower traces. The black lines show the fit for exponentials. }
\end{figure}

\begin{figure}[ht!]
		\centering
		\includegraphics[width=1.0\textwidth]{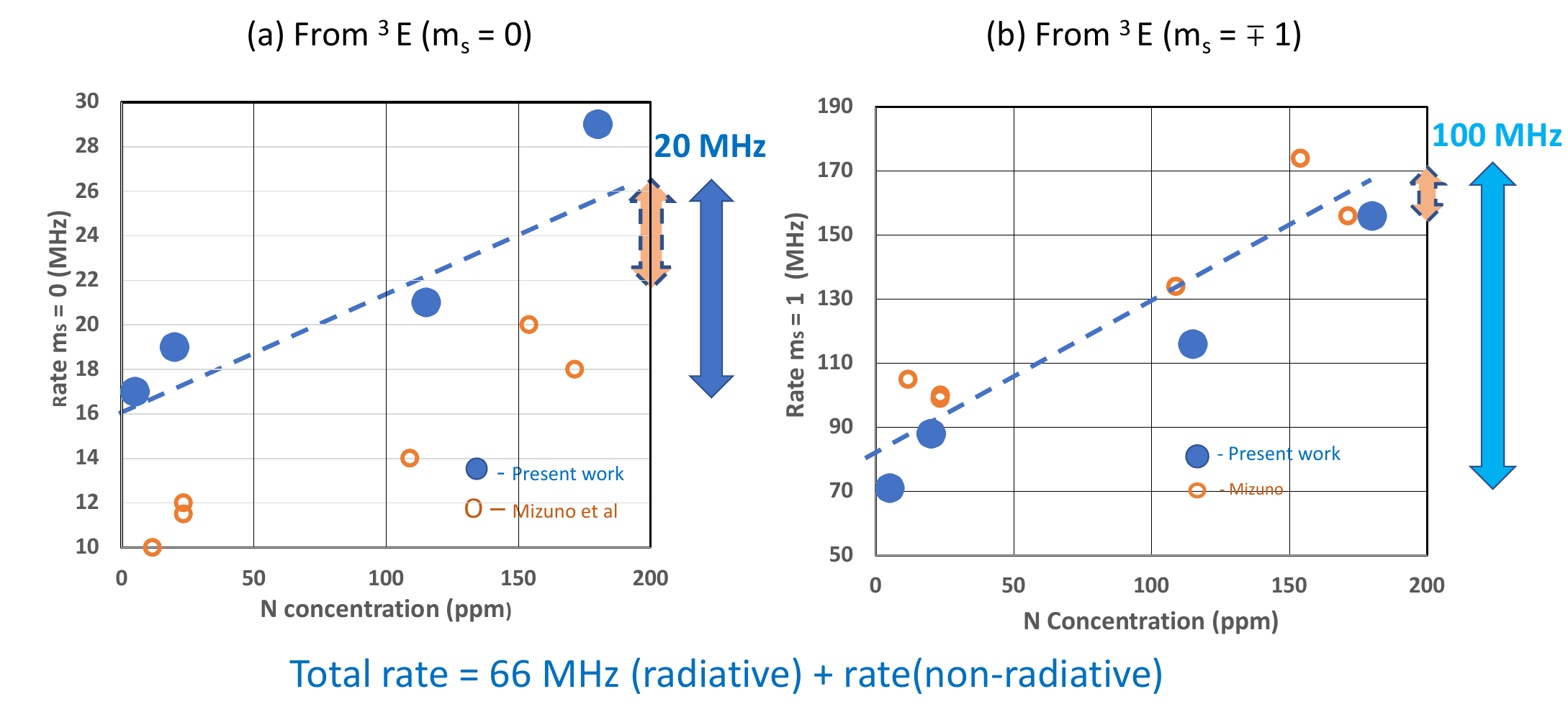}
		\caption{\label{Maybe lifetimes}  Total decay from NV$^-$  $^3$E state is due to (i) the radiative decay of 66 MHz plus non-radiative decay from cross relaxation to singlet and from charge transfer to NV$^0$. The graph give the value of the  non-radiative components. The blue circle is for the present data and the orange circles are similar measurements taken from the figures of  Mizuno \textit{et al.} \cite{Mizuno2021} The lifetimes in the left figure are for the m$_s$ = 0 spin component for N-concentrations from 0 ppm to 200 ppm. With this range of concentration the rates vary from component 17 MHz (12 ns) to 29 MHz(10.5 ns). The figure on the right gives the related rates for the m$_s$ = $\pm$1 spin component. They are significantly faster varying from 71 MHz (7.3 ns) to 156 MHz (4.5 ns). The dashed line is only a guide to the eye. The contrast between the two spin states is highlighted by the blue arrows on the right. The change arises for both intersystem crossing and charge transfer. The contribution from charge transfer must be the same for both spin states as the process is shown to be spin independent. Therefore, the approximate magnitude of the contribution due to charge transfer is indicated (but not quantitative) by the orange arrows as it must be less than 12 MHz. }
\end{figure}

The lifetimes were measured using a 488 nm and/or a 635 nm pico-second laser and emission responses were detected with a Si avalanche photo diode. To  avoid including the detection of NV$^0$ response the emission was restricted to long wavelengths > 750 nm. The measurements were made at room temperature with no spin polarisation and equal population in each of the three ground state spin levels. The traces were, therefore, fitted with two exponentials of equal weight. With this functional form it was possible to obtain satisfactory agreement with experiment as shown in Fig \ref{lifetimes}. The values for the slower decays corresponding to that from the m$_s$ = 0 spin levels and the faster from m$_s$ = $\pm$1 levels. Representative experimental observation for four samples are shown in Fig \ref{Maybe lifetimes} with the intention of highlighting the change of the lifetimes with nitrogen concentration. Values of the six samples given earlier in Fig \ref{samples} and are also given in Table \ref{table}.
	
In Fig \ref{Maybe lifetimes} and in Table \ref{table} four values are given associated with each measurement. The lifetime for each fit is given in ns and the corresponding rate in MHz: slow for  m$_s$=0 and fast for m$_s$=$\pm$1. These rates arise from a radiative and non-radiative component. The radiative rate is 66 MHz and what varies is the non-radiative component. It is the non-radiative component that is included in Table \ref{table} as a+J and B+J in MHz. For example, the shortest lifetime for the 5 ppm sample of 12 ns for the m$_s$ = 0 state implies a total decay rate of 83 MHz. This is due to 66 MHz radiative decay plus non-radiative rate  of 17 MHz and it is this latter value that is given in Table \ref{table} and plotted in Fig \ref{Maybe lifetimes}. The non-radiative rates for m$_s$ = 0 gives an overall change of order of 20 MHz for nitrogen concentrations varying from 5 to 181 ppm (left plot in Fig \ref{Maybe lifetimes}). This contrasts with the faster rates for m$_s$ = $\pm1$ where the change is of order of 100 MHz. In Table \ref{table} and Fig \ref{Maybe lifetimes} it is noted that the same trend was reported by Mizuno \textit{et al.} \cite{Mizuno2021} for oriented samples. At the top of the Table \ref{table} there are six values for single sites that have been reported and these are included for comparison. 	
What is also given in Table \ref{table} is the values obtained earlier for the ratio of the singlet decay relative to the visible decay where spin polarisation is quenched, and the ratio does not involve J charge transfer. This is considered to give the value of (a +2B)/3. In the case of single site (top six values) and Mizuno samples the (a + 2B)/3 values are calculated from lifetimes. 

\begin{table}[]
	\resizebox{\textwidth}{!}{\begin{tabular}{llllllllll}
	\toprule
	 Reference & year & name    & N conc                   & na         & a+J (MHz)                  & ns         & B+J (MHz)                  & a +2B/3                &  \\
			\midrule
			Batalov     & 2008 & single  & single                   & 12 ns                    & 0.25 (17MHz)             & 7.8 ns                   & 0.94 (62 MHz)            & 0.71                     &  \\
			Robledo     & 2011 & single  & single                   & 13.7ns                   & 0.11 (7 MHz)             & 7.3 ns                   & 1.07 (71 MHz)            & 0.75                     &  \\
			Sturmer     & 2012 & single  & single                   & 12.2 ns                  & 0.24  (16 MHz)           & 8.6 ns                   & 0.75 (50 MHz)            & 0.58                     &  \\
			Tittiene  & 2012 & Average & single                   & 13.7 ns                  & 0.11 (7 MHz)             & 8.6 ns                   & 0.75 (50 MHz)            & 0.54                     &  \\
			Goldman     & 2015 & single  & single                   & 12 ns                    & 0.26 (17 MHz)            & 7.5 ns                   & 1.0  (67 MHz)            & 0.71                     &  \\
			Gupta       & 2016 & Average & single                   & 12.9 ns                  & 0.17 (11.5 MHz)          & 6.3 ns                   & 1.4 (93 MHz)             & 0.99                     &  \\
			&   --&       --  &                 --      &                     --   &                     --   &          --                &                  --        &              --            &  \\
			&      & DNV     & 0.8 ppm                  & 13 ns                    &                          & 7 ns                     &                          &                          &  \\
			&      & LAD     & 5ppm                     & 12ns                    & 0.26 (17 MHz)            & 7.3 ns                   & 1.1 (71MHz)            & 0.82 (0.61)                    &  \\
			&      &         &                          &                          &                          &                          &                          &                          &  \\
			Mizuno      & 2021 & 6 . 2   & 11.7 ppm                 & 13.1 ns                  & 0.15 (10 MHz)            & 5.85 ns                  & 1.59 (105 MHz)           & 1.11                     &  \\
			
			Mizuno      & 2021 & 8  . 3  & 23.7 ppm                 & 12.8 ns                  & 0.18 (12 MHz)            & 6 ns                     & 1.53 (101 MHz)           & 1.06                     &  \\
			Mizuno      & 2021 & 6  . 1  & 23.4 ppm                 & 12.9 ns                  & 0.17 (11.5 MHz)          & 6.1 ns                   & 1.48 (98 MHz)            & 1.04                     &  \\
			&      & FOX     & 25 ppm                   & 11.7 ns                  & 0.29 (19 MHz)            & 6.5 ns                     & 1.3 (88MHz)            & 0.96 (0.6)                 &  \\
			Mizuno      & 2021 & 8 . 1   & 109 ppm                  & 12.5 ns                  & 0.21 (14 MHz)            & 5 ns                     & 2.03 (134 MHz)           & 1.4                      &  \\
			&      & RAN     & 115 ppm                  & 11.5 ns                  & 0.32 (21MHz)            & 5.5 ns
			             & 1.75 (116 MHz)                         & 1.27 (0.84)                       &                                        \\
			Mizuno      & 2021 & 3 . 2   & 154 ppm                  & 11.6 ns                  & 0.33 (22 MHz)            & 4.2 ns                   & 2.6 (172 MHz)            & 1.8                      &  \\
			Mizuno      & 2021 & 3 . 4   & 171 ppm                  & 11.9 ns                  & 0.27 (18 MHz)            & 4.5 ns                   & 2.4 (156 MHz)            &  1.7                        &  \\
			&      & RAB     & 181 ppm                  & 10.5 ns                  & 0.44 (29 MHz)            & 4.5 ns                   & 2.4 (156 MHz)            &1.74 (0.95)                    &  \\
			&      & TRI     & 600 ppm                  & 6 ns                     & 1.5 (100 MHz)           & 1.7 ns                   & 7.9 (522 MHz)            & 5.8   (1.0)    &  \\
			\bottomrule 
	\end{tabular}}
	\caption{\label{table}
		Parameters for NV$^-$ centre for single site (upper) and samples containing nitrogen (lower). The upper six set of values give the lifetimes of single NV$^-$ centres reported in the literature by Batalov \textit{et al.}\cite{Batalov2008}, Robledo \textit{et al.}\cite{Robledo2011}, Sturner \textit{et al.} \cite{Sturner2021}, Tetienne \textit{et al.}\cite{Tetienne2012}, Goldman \textit{et al.}\cite{Goldman2015a} and Gupta \textit{et al}\cite{Gupta2016}. The single site values display considerable variation, but this is not discussed in this paper. The lower values are for the NV$^-$ where the sample contains a significant concentration of single substitutional nitrogen. The lifetimes are from Mizuno \textit{et al}. \cite{Mizuno2021} and this work in capitals. Mizuno~\textit{et al.} give lifetimes but also report spin polarisation from EPR measurements. The last column gives the value of a+2B/3 calculated from lifetimes. The numbers in brackets are from experimental visible/IR ratios and it is noted that the values are consistently lower than that predicted from lifetimes.}
\end{table}

\subsection{Spin polarisation and contrast: Discussion  of change of optical cycle}
The lifetimes indicate an increase of intersystem crossing with nitrogen, but the origin of this increase is not clear. The increase is unlikely to be due to the electric fields associated with adjacent N$^+$ donors as there is no change when the fields are increased when moving the donors closer by illuminating with red light as shown in reference \cite{Manson2018}. Also, although coupling triplets and singlets to the spin 1/2 of nitrogen atoms can enable a crossing without spin change, this certainly is not the mechanism as it does not occur for the lower intersystem crossing. The lower singlet-triplet crossing is the equivalent situation but it has been shown that there is no change with nitrogen concentration (Fig \ref{singlet E}). An alternative consideration is that the increase originates from a  change of the local vibration distribution in the crystals due to nitrogen.  Goldman\textit{et al.} \cite{Goldman2015a,Goldman2015b} has shown that the spin-orbit interaction enabling the upper intersystem crossing from $^3$E to $^1$A$_1$ involves an energy transfer with a particular vibrational frequency and so a change of vibrations could alter the crossing. There is no obvious difference observed for the visible or infrared vibrational sidebands between the lowest and highest concentrations although it is known that there are changes at very low frequency as infrared absorption is used to give a measure of the concentration of substitutional nitrogen. Hence this latter mechanism cannot be ruled out but the reason for the change of the intersystem crossing remains a outstanding issue.

There is an interest in the increase of the crossing rates as it is the rates that determine the spin polarisation. An improvement in polarisation requires changes to the a/B ratio but here both a and B increase with nitrogen with the ratio near constant. As far as the authors are aware the origin of the m$_s$= 0 rates (a-parameter) are not so well established as that for m$_s$ = $\pm$1. (B-parameter). As in the previous paragraph the reason for the increases are not understood. The near fixed a/B ratio indicate there might be a mixing of the spin state but the cause of any such mixing is likewise not known. It does leave the intriguing possibility that it may be possible to change the ratio and improve spin polarisation.

\section{Spin quantum state and spin-flip transtions}
\begin{figure}
	\centering
	\includegraphics[width=1.0\textwidth]{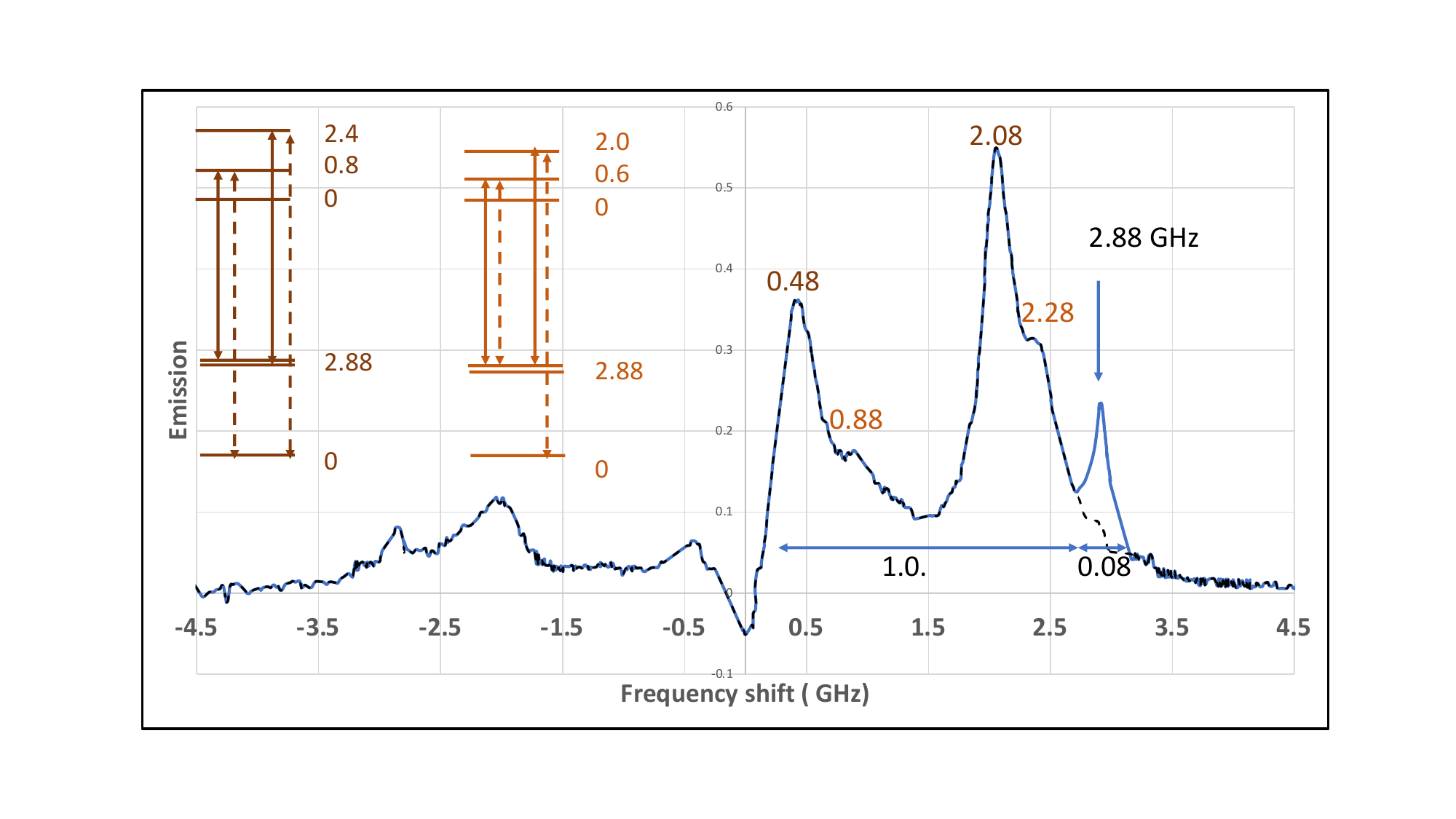}
	\caption{\label{hole}. Spectra of two laser holeburning of NV diamond. One laser is fixed in frequency at a wavelength within the NV$^-$ zero-phonon line. This laser is switched off and a second laser is swept in frequency about that of the fixed laser \cite{Reddy1987}. Values are from Reference \cite{Manson1994}. The area under the feature at 2.88GHz is 5.0 $\pm$0.5 percent of the area under the total of the positive features. }
\end{figure}

In  early hole-burning measurements of the NV centre  gave transitions where spin was not conserved. Transitions were observed at frequency shifts of 2.88 GHz as shown in Fig \ref{hole}. This frequency within NV$^-$ was known to be associated with a splitting of $S=1$ spin system and the  observation of an absorption (anti-hole) established that the spin was within the ground state. The other features in the spectrum in Fig \ref{hole} involve spin conserving transitions between split levels of the $^3$E orbital doublet and the ground state. The sample was a 1b with nitrogen and the splitting arises from the electric fields from N$^+$ donors  \cite{Block2021,Manson2018} and the magnitude can be several 100's GHz significantly larger that D$_g$ and D$_e$. Each of the split components of the orbital doublet have three spin levels and where the orbital splittings are large the separation between m$_s$ = 0 and the other spin projections in both branches approach values of 3D$_{2A_1}$  $\pm$ D$_{2E}$ (=0.645 GHz and 2.145 GHz)  where D$_{2A1}$ and D$_{2E}$ are spin-spin interactions (see Figure 4 in reference \cite{Doherty2011}). With contribution from spin-orbit interaction these separations can differ in the two branches and here the splittings are 2.0 GHz and 0.6 GHz ( in lower branch ) and 2.4 GHz and 0.8 GHz (in higher branch) \cite{Reddy1987,Manson1994} and these values give rise to the broad features in the spectrum in Fig \ref{hole}. More particular with m$_s$ = 0 and m$_s$ = $\pm$1 spin levels being close (particularly the 0.6 GHz in the lower branch) spin-spin interaction causes a mixing of the spin states and result is the sharp non-spin-conserving transition to the ground state at 2.88 GHz.  There is inhomogeneities but not large owing to the asymptotic behavior of the spin separations. The result is that the mixing and characteristics of the  hole-burning spectrum are fairly consistent. The conclusion is that spin-flip transitions as observed in the hole-burning experiments are typical of NV ensembles in crystals containing nitrogen. Where experiments involve this situation calculation for low temperatures need only include additional transition about 5 percent, of the conserving transition strengths to allow for these spin-flip transitions.

\begin{figure}
	\centering
	\includegraphics[width=1.0\textwidth]{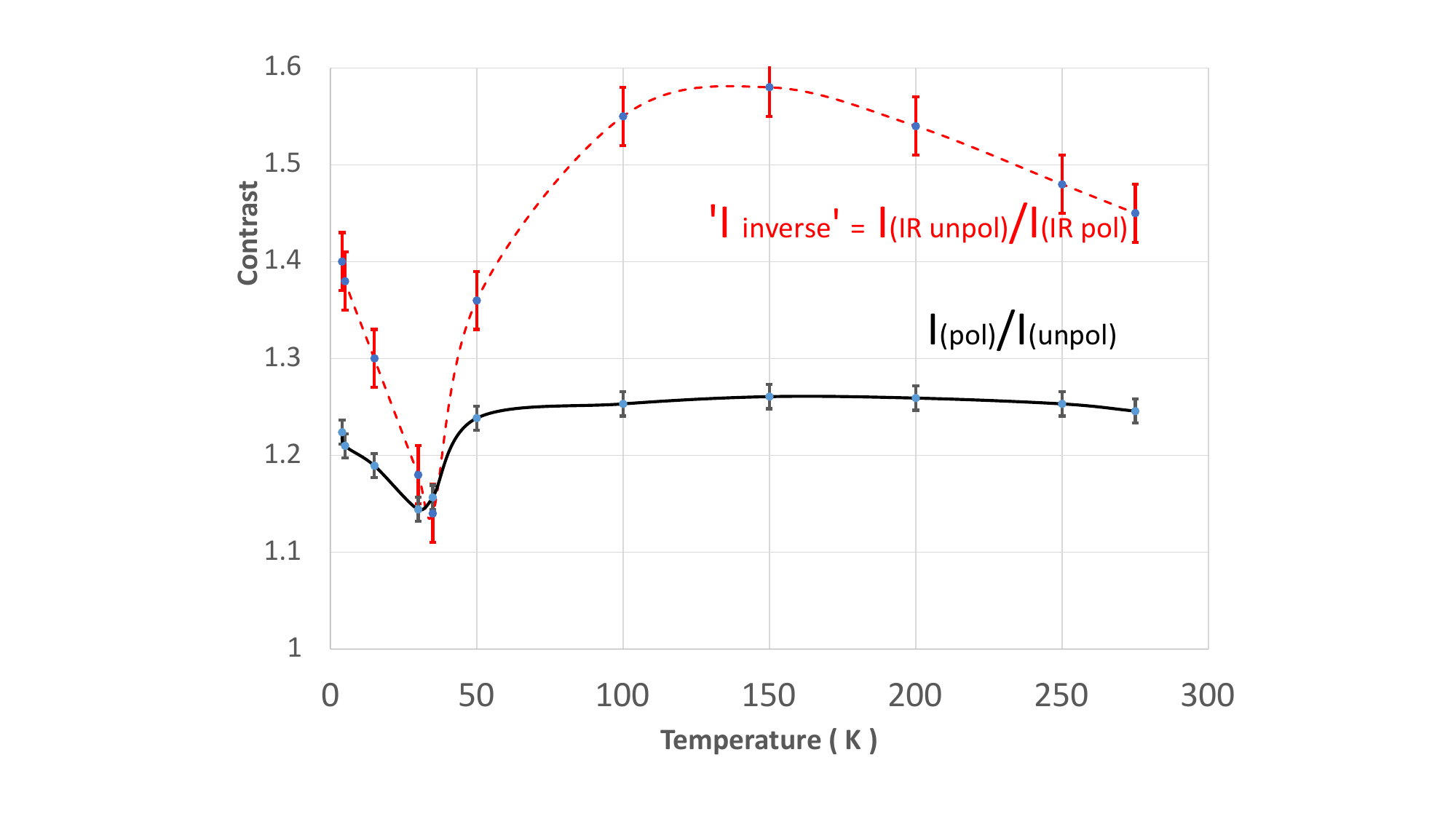}
	\caption{\label{electric}
		Temperature dependence of visible and infrared contrast for the ensemble sample as used for holeburning spectrum (similar to 181 ppm sample). The visible contrast is given as emission/unpolarised emission with unity equivalent to unpolarised. The infrared response is reversed.  At 77K (or room temperature) when spin polarised the visible emission \textit{increases} by 25 per cent  (from 1, unpolarised to 1.25, polarised) and IR \textit{decreases} by 27 percent (from 1 to 0.63). The upper trace is the inverse and is included to highlight the larger fractional change of the infrared response -- also obvious in the following Fig \ref{fourspectra}. The reduction of emission above 150 K could be due to having not integrating emission over a sufficiently large spectral region with the broadening with temperature.}
\end{figure}

The situation for single centres at low temperature (4K) is different to that given above. The $^3$E energy levels of single centres are dependent on strain and electric field but the large splittings as above do not occur. The spin levels can be much closer and as a result give larger mixing. The most significant case is where there is avoided crossing of the levels and almost `total' mixing. The spin flip transitions can then be observed in absorption \cite{Tamarat2008}. There is not a treatment of single centres that fit all cases and at low temperatures each centre has to be treated separately. The situation changes with increasing temperature due to phonon-mediated state averaging. This process was first suggested for ensemble samples where a decrease of NV emission at 30K was noted by Rogers \textit{et al}.\cite{Rogers2009} (only shown as in insert in Fig 4 of that publication) and equivalent observations are given here in Fig \ref{electric} and the variation of the infrared emission is added. The reason for the decrease in emission was not well understood but recently there have been a complete treatment of phonon-mediated state averaging for single sites by Happacher~\textit{et al.} \cite{Happacher2022,Happacher2023} and also by Ernst \textit{et al}\cite{Ernst2023a,Ernst2023b}. They have shown that the intensity variation is associated with the phonon-mediated jumping between the two orbital branches. With optical excitation there is spin polarisation but as described above there is spin mixing that reduces the emission. The mixing is predominantly in the lower of the two orbital branches of the excited state. As temperature is increased there is some jumping between the branches and this initially (at 50 K for single site ) has the effect of increasing the mixing and reducing the emission. With further increase of temperature, the jump rates increase and become faster than the spin-mixing process such that the mixing is reduced. The spin polarisation improves and the emission intensity increases as seen in the case of an ensemble in Fig \ref{electric}. It is general considered that by a temperature of 150 K the rates are such that spin mixing is negligible and spin becomes a good quantum number and certainly this is the situation at room temperature. As a result at room temperature throughout the literature the spin in the excited state is consider a good quantum number and excited state is treated as an effective orbital singlet similar to the ground state

Happacher \textit{et al.} \cite{Happacher2022} illustrates that for a single site with a small orbital state splitting (eg 1.7 GHz) the emission minimum occurs at 50 K with an effective full width of the order of 20 K although slow `recovery' towards higher temperatures. With higher strain splitting, approaching 100 GHz splitting, the emission minimum is reduced in temperature to $\sim$30 K but is broader such that there is significant reduction of spin polarisation even at zero K. The changes with orbital splitting are largely due to the variation of the energy separation of spin states and the latter is more akin to that reported above for ensembles. Indeed for ensembles the orbital splitting ranges from 100 GHz to 1000 GHz as can be seen from the optical zero-phonon line width \cite{Block2021}. There is no change to the processes involved within this range but for the ensemble one must allow for this distribution. The results for emission temperature are that the minimum is at 30 K but broad as shown Fig \ref{electric}. Above 30 K the spin polarisation and emission increases with a maximum at 150 K. This variation is clearer in the infrared response although the signals size is very much smaller. The `recovery' is not complete by 77 K and this indicates that there could be some spin mixing at 77K. This does raise the issue that there is not a simple way of establishing the presence of spin-flip transitions at temperatures where spectral hole-burning are not informative. Fortunately, any mixing at 77K will be smaller than other uncertainties and not generally included and certainly not elsewhere in this paper.

The conclusion is that for ensembles containing nitrogen involving large splitting (>100 GHz) of the orbital doublet the state averaging results in negligible spin mixing above 100 K making ensembles suitable for most application including magnetometry above this temperature.  Surprisingly this is not the case for single sites but largely because of small splitting of the excited state. The m$_s$ =0 and m$_s$ = $\pm$1 spin levels are always close (< 10GHz) and  spin-mixing is more prevalent. This is particularly the case when magnetic fields are involved as there are situations where particular fields can be precisely at the value giving an avoided crossing of the spin levels and very large mixing. Consequently, there are temperatures and fields where there is only weak spin polarisation and the emission change with microwaves can be very small. The single sites enable great spactial resolution but at the low temperatures ($\ll100$~K) the response for magnetometry as just described is irregular. There will be ways round this problem but regardless the situation is totally changed above 150 K as the state averaging eliminates all the complications of excited state level crossing and there is excellent magnetic sensitivity as well as great spatial resolution as reported by many.

\section{Summary and correspondence with experiment}
\begin{figure}[ht!]
  \centering
    \begin{subfigure}[b]{0.95\textwidth}
        \includegraphics[width=\textwidth]{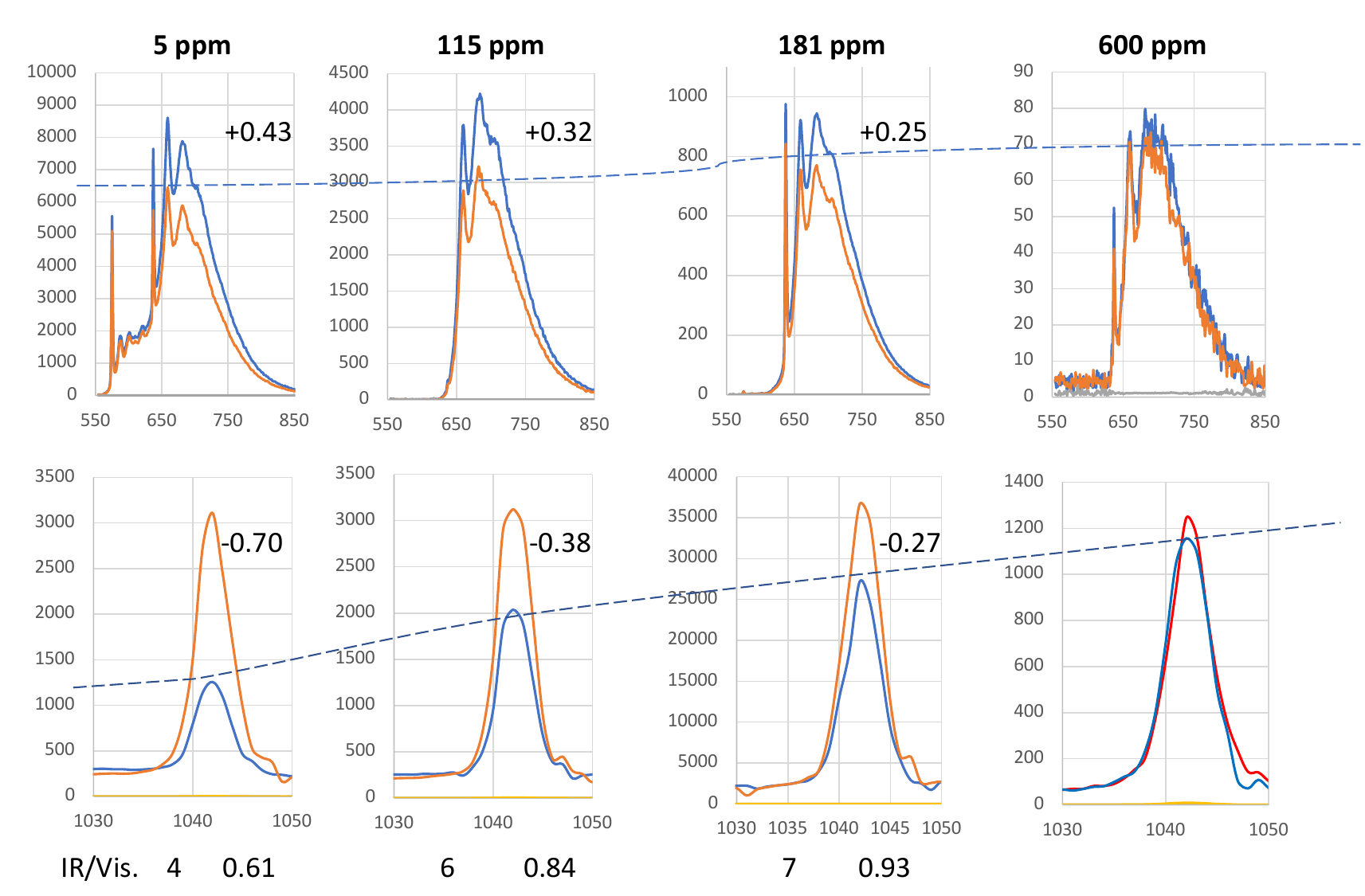}
        \caption{}
        \label{}
    \end{subfigure}

    \begin{subfigure}[b]{0.31\textwidth}
        \includegraphics[width=\textwidth]{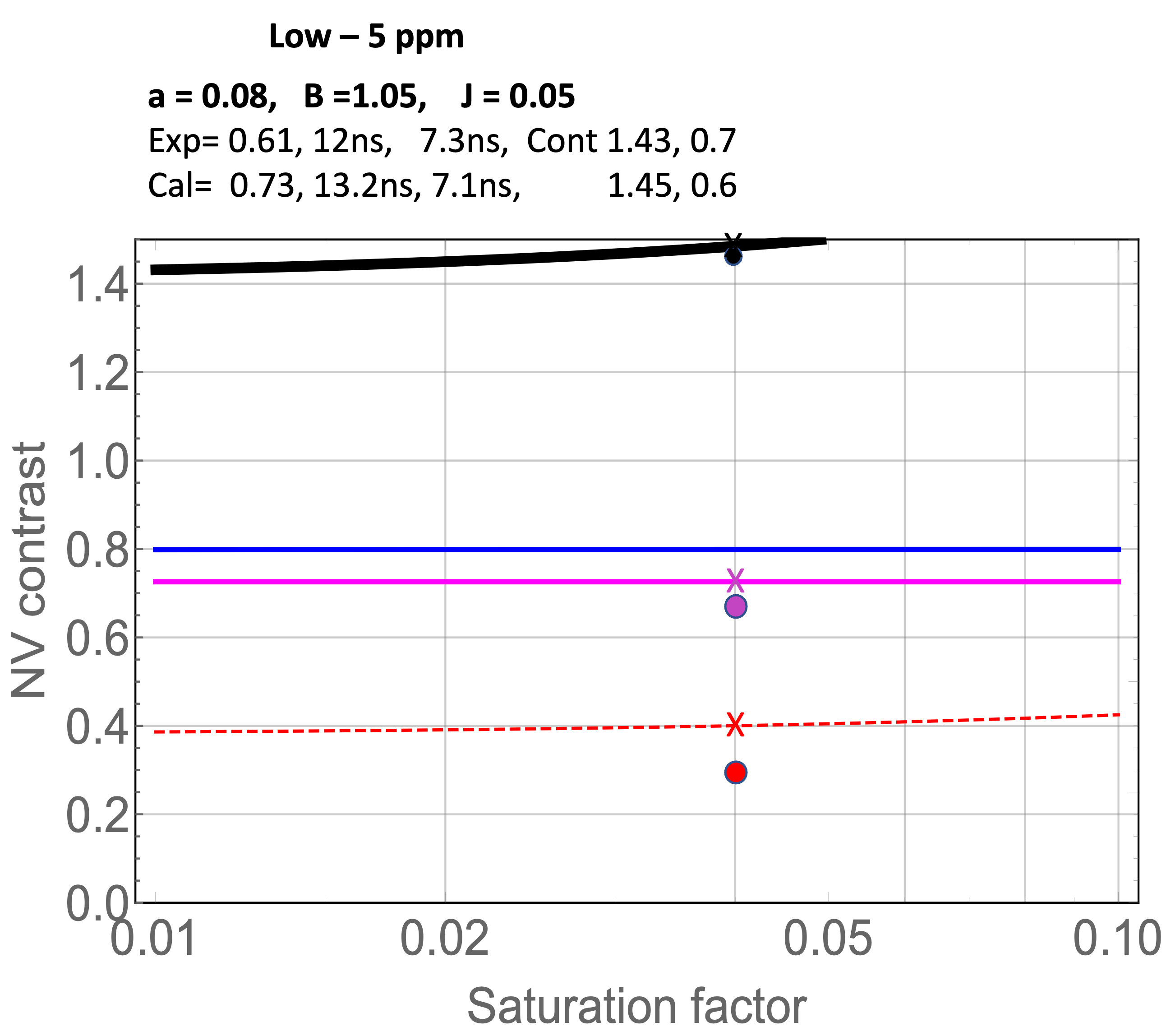}
    \end{subfigure}
~
    \begin{subfigure}[b]{0.31\textwidth}
        \includegraphics[width=\textwidth]{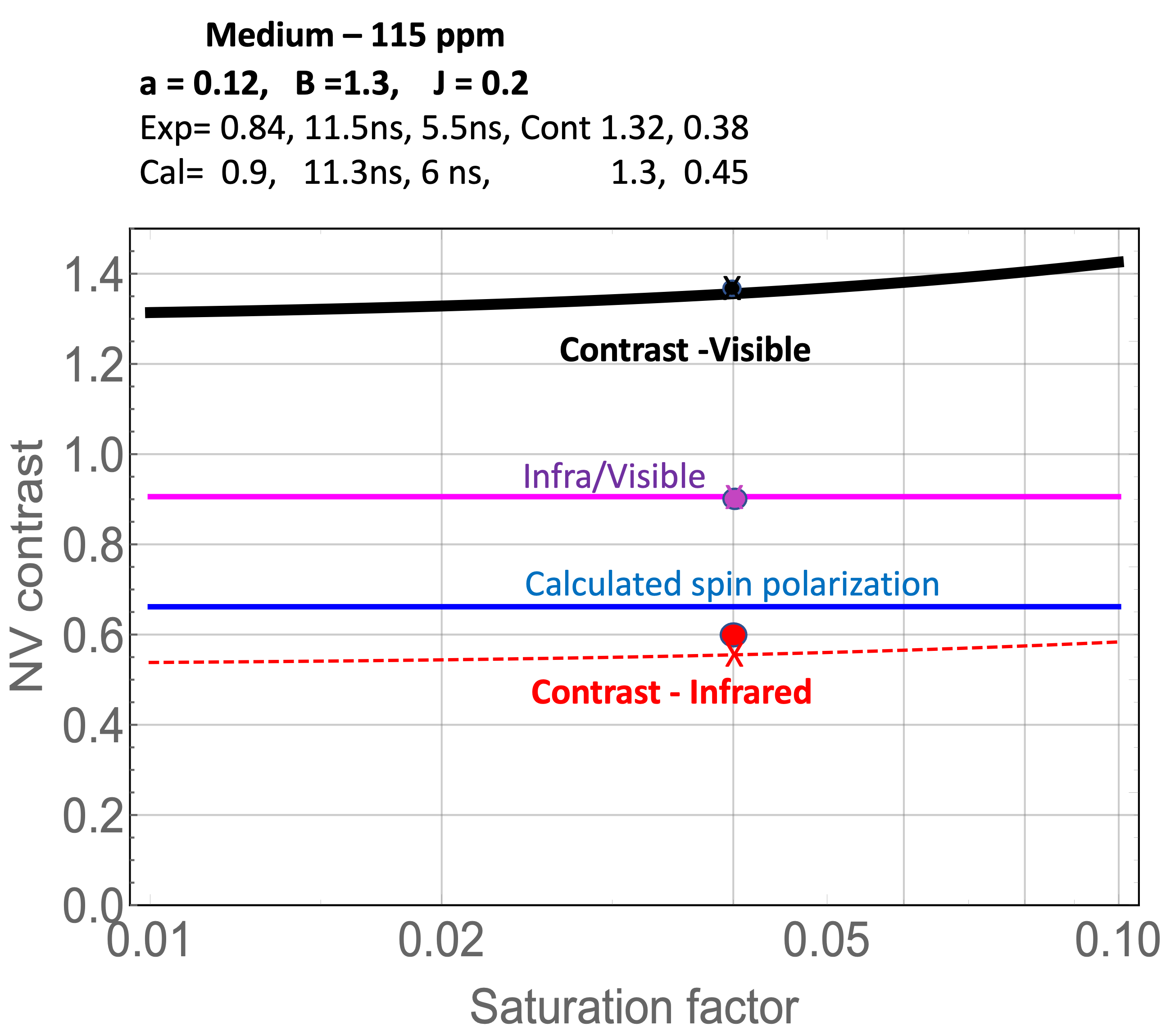}
    \end{subfigure}
~
    \begin{subfigure}[b]{0.31\textwidth}
        \includegraphics[width=\textwidth]{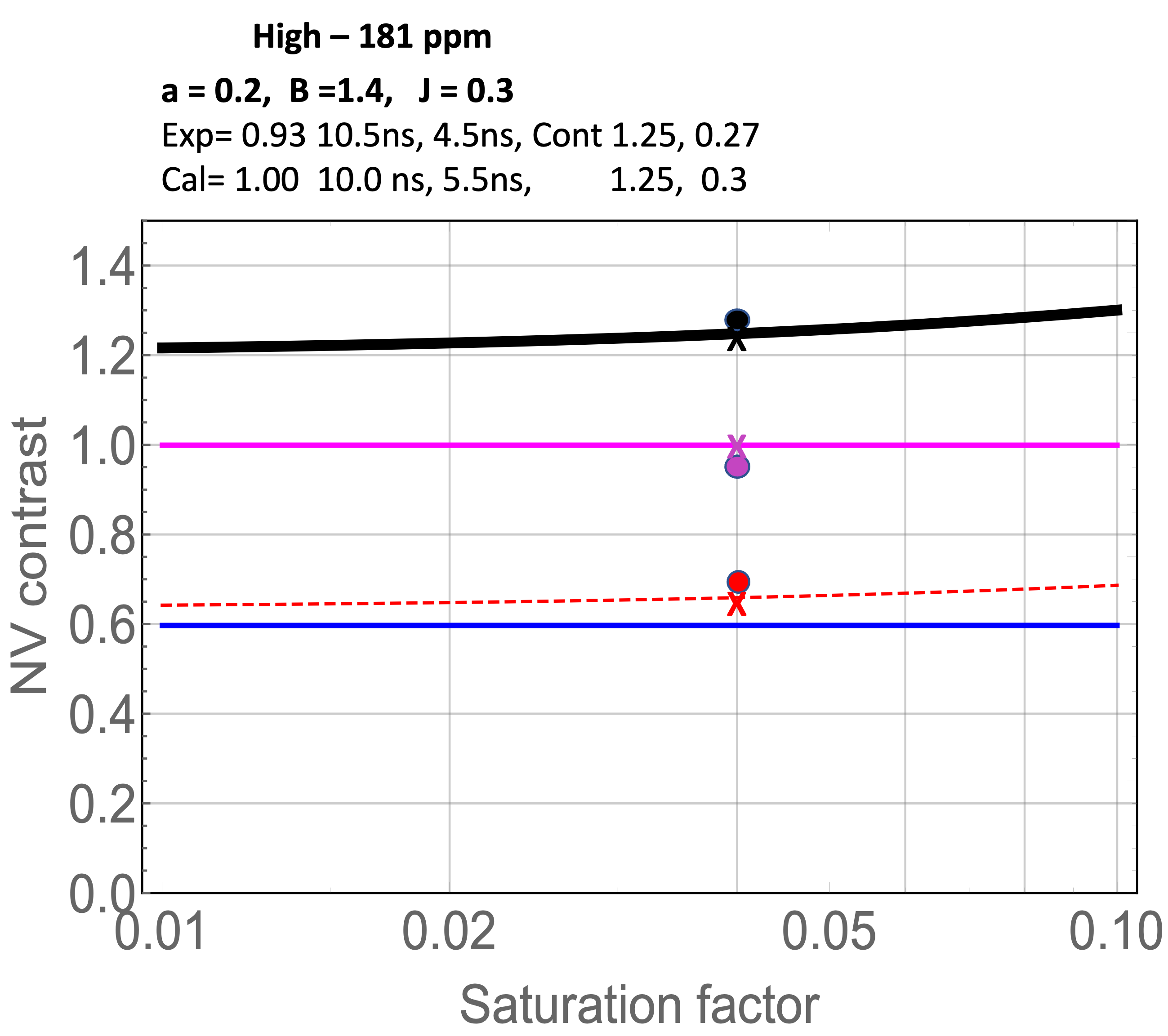}
    \end{subfigure}
    \caption*{(b)}
	\caption{\small \label{fourspectra}  (a) To highlight the trends with nitrogen concentration details are given for four samples from  5~ppm to~600 ppm. (other samples have been given in Fig \ref{samples} but in less detail). Three samples correspond to the Low nitrogen (5 ppm), Medium nitrogen (115 ppm) and High nitrogen (181 ppm). The fourth spectra here corresponds to that for very high nitrogen (600 ppm) for which there is negligible polarization and parameters have not been determined. In the upper spectra the polarised emission (no field) in blue are normalised and the unpolarised spectra with field are in red. The dashed line as guide to the eye indicates the change of contrast with increasing nitrogen. Almost the reverse for the infra-red in the lower traces where unpolarized spectra in red are normalised and the polarized in blue increases with nitrogen concentration. In the lower Figures (b) by adjusting of the three parameters  a, 2B and J plausible correspondence is obtained for the data for the three samples above. (For the modelling of the NV$^-$ emission.  There is no calculation for the very high nitrogen sample. The black and dashed red lines give the changing contrasts for visible and infrared emission respectively. Magenta indicates the experimenally and calculated values of the ratio of the decay via the singlet in relation the visible emission. This ratio is also given in the first number above. Lifetimes for m$_s$ = 0 and m$_s$= 1 states, contrast  visible and infrared both experimentally and calculated using above parameters are also given. The calculated values are given by crosses which should be compared with experimental measurements given as circles. Calculation of the spin polarisation are given in the figures in blue but there is no experimental measurement.}
\end{figure}

A summary of the trend of the contrast for the 77K visible and infrared are shown in Fig \ref{fourspectra}(a) for three samples (at nitrogen temperatures) low (5 ppm), medium (115ppm) and high(181ppm) nitrogen concentrations. The calculation in Fig \ref{fourspectra}(b) give plausible correspondence with the spectra. Parameters are largely determined from lifetimes with contribution from inter-system rates (a and B) and smaller contribution from charge transfer (J).  The excitation in all cases is low and there is no ionisation of the nitrogen. The three parameters a, B and J are increased marginally for increasing nitrogen concentration. The laser does excite NV$^0$ and so transition K (Fig \ref{pair}) is included but to first order the magnitude does not affect the calculated values. In the calculation the lower intersystem rates are taken from the single site literature as in Table \ref{table} c= 0.04 (2.6 MHz ) and D = 0.08 (5.2 MHz). Here it is considered that this crossing does not alter the spin polarisation \cite{Robledo2011} although it is noted that more recent experiments have indicated that a higher ratio of 1: 1: 8 might be more appropriate than 1: 1: 2 \cite{Kalb2018}. Overall, the agreement with experiment obtained in Fig \ref{fourspectra} is reasonable. The agreement could be improved with further adjustment of the parameters but with as many parameters as observation will not be meaningful. More significantly it is recognised that each sample contains centres with distribution of properties and using `one set of parameters' for each nitrogen concentration a reasonable correspondence with experiment is taken to indicate that the processes have been adequately identified and satisfactory average values adopted.

\section{Discussion }
This paper starts by highlighting that the fact that for the NV$^-$ centre involving diamonds containing nitrogen the properties are different from those of single centres in high purity diamond. It is questionable as to how broadly this difference is recognised but there certainly seem to be confusion with charge transfer. With nitrogen in the lattice there is charge transfer due to electron tunnelling between nitrogen and the NV centre and this does not occur with single site. This phenomenon has been reported previously but the treatment here is more extensive. Fig \ref{pair} summarises the situation and contrasts with the more documented case of single NV ionisation of Fig \ref{single} as there is no involvement of valence or conduction electrons. With nitrogen three tunnelling transitions are considered in detail, two NV$^0$ $\rightarrow$ NV$^-$ and one NV$^-$ $\rightarrow$ NV$^0$. These cases are known previously but rates are shown to dependent on nitrogen concentration and the transition processes concluded to be spontaneous. The transitions can combine to give cycling of the NV charge state and the cycling reduces the spin polarisation. The transitions can shorten the optical lifetimes, but the lifetimes are more effected by modifications of the spin-triplet/singlet intersystem crossing.  The modification of the intersystem crossing has not been reported previously. As well as giving significant changes to the lifetime the different intersystem crossing also increases the infrared emission and the decay via the singlets making the visible emission `less bright'. When allowing for both the modification of the intersystem crossing and the tunneling rates a simple five level rate equation gives reasonable correspondence with the observed trends of optical contrast, infrared contrast and lifetimes. The correspondence between experiment and theoretical model is encouraging in that the major process and the systematic variation that effect the degrading of the NV properties have been identified. No attempt is made at a first order calculation of the magnitude of the changes, and this is left as an outstanding issue.

Aspects of charge transfer via tunneling determines the approach for many ensemble experiments. For example, it might be possible to use a red laser to excite NV$^-$ centre as this would result in spin polarisation but the complication is that if only $^3$A$_2$ $\rightarrow$ $^3$ E is excited there is a continuous loss of signal. The loss is a due to the NV$^-$ $\rightarrow$ NV$^0$ tunneling, here termed transition L. To avoid this loss the convention is to use 532 nm (2.33 eV or higher energy) as the laser then also excites NV$^0$ and introduces the reverse NV$^0$ $\rightarrow$ NV$^-$ tunneling, termed J, that stabilises the emission. However, there is still a difficulty as at this wavelength the laser can ionise the nitrogen bath and lead to an alternative but equivalent NV$^-$ $\rightarrow$ NV$^0$ charge transfer. There is a loss of NV$^-$ concentration and with cycling loss of spin polarisation. Furthermore, if NV$^0$ emission is included in the detection the overall contrast would be reduced and so emission of NV$^0$ has to be excluded. This involves restricting the optical band to $>700$ nm)  and results in a further loss of signal.

By using 575 nm excitation some of the latter difficulties can be avoided and this is illustrated using a dye laser at this wavelength. The laser excites NV$^-$ to created spin polarisation and emission. It also excites NV$^0$ and again give the  NV$^0$ to NV$^-$ charge transfer to stabilise the emission. However, at this energy of 2.15 eV the ionisation of nitrogen is almost negligible and there is not the loss of NV$^-$ via one of the NV$^-$ $\rightarrow$ NV$^0$ charge transfer mechanisms. There is still some charge transfer but only from J tunnelling and in this case, the NV$^0$ created has the same spin contrast as NV$^-$. Consequently, with same contrast there is no need to reduce the optical detection bandwidth. Both NV$^0$ and NV$^-$ emission can be detected giving a second clear advantage of using 575 nm rather than 532 nm. In short with 575 nm there is no nitrogen ionisation and all emission can be detected. Of course, to date there is no simple commercial laser at this wavelength, but this might change.

In summary the aim of the work has been to highlight the main processes that degrade the properties of NV when there is nitrogen in the sample and this aim has been largely achieved. Repeating the degrading processes are the changes to charge transfer, changes to the intersystem crossing and for green laser excitation ionisation of the nitrogen bath. It is also reminded that spin-flip or  non-spin-conserving transitions do occur and are of indetermined magnitude between 0 K and 77 K but very small at liquid nitrogen temperature and zero at room temperature.

\section{Acknowledgements}
Thanks to Ronald Ulbricht for many discussions about the NV centre.
Also thanks to Marco Capelli and Phillip Reineck for sharing their lifetime measurements.
Thanks to Bryce Henson for assistance with the Swabian Time Tagger for the measurements of optical  lifetimes.
Also thanks for the many interactions with colleagues, Matt Sellars, and Rose Ahlefeldt.
N.B.M. thanks the Australian Research Council  for grant DP 170102232. 
M. W. D. is indebited to  Australian Research Council for the award of DE 170100169. 
\section{ORCID iDs}
		
Neil B Manson https://orcid.org/0000-0002-5875-4118,
Morgan Hedges https://orcid.org/0000-0001-9714-7356,
Michael S J Barson  https://orcid.org/0000-0003-0247-5619,
Carlos Meriles https://orcid.org /0000-0003-2197-1474,
Marcus W Doherty https://orcid.org/0000-0002-5473-6481,

\bibliography{document.bib}
\bibliographystyle{unsrt}
\end{document}